\documentclass[aps,showpacs,nofootinbib]{revtex4}
\usepackage{epsfig}
\usepackage{graphicx}
\usepackage{amsmath,amstext,amsthm,amssymb}
\usepackage{wrapfig}

\begin{document}

\title{Multiplicities in
Au-Au and Cu-Cu collisions at $\sqrt{s_{NN}} = 62.4$ and $200$ GeV}
\author{Dariusz Prorok}
\affiliation{Institute of Theoretical Physics, University of
Wroc{\l}aw,\\ Pl.Maksa Borna 9, 50-204  Wroc{\l}aw, Poland}
\date{September 12, 2013}

\begin{abstract} Likelihood ratio tests are performed for the hypothesis that
charged-particle multiplicities measured in Au-Au and Cu-Cu
collisions at $\sqrt{s_{NN}} = 62.4$ and 200 GeV are distributed
according to the negative binomial form. Results suggest that the
hypothesis should be rejected in the all classes of collision systems and centralities of PHENIX-RHIC measurements. However, the application of the least-squares test statistic with systematic errors included shows that for the collision system Au-Au at $\sqrt{s_{NN}} = 62.4$ GeV the hypothesis could not be rejected in general.
\end{abstract}

 \maketitle


\section{Introduction}
\label{intro}

The analysis of charged hadron multiplicities in Au-Au and Cu-Cu
collisions at $\sqrt{s_{NN}} = 62.4$ and 200 GeV was done by the
PHENIX Collaboration in \cite{Adare:2008ns}. It was also claimed
there that these multiplicities are distributed according to the
negative binomial form. The UA5 Collaboration noticed for the first
time that charged-particle multiplicity distributions measured in
high energy proton-(anti)proton collisions in limited intervals of
pseudo-rapidity have this form \cite{Alner:1985zc,Ansorge:1988kn}.

The Negative Binomial Distribution (NBD) is defined as

\begin{equation}
P(n; p, k) = \frac{k(k+1)(k+2)...(k+n-1)}{n!} (1-p)^{n}p^{k} \;,
\label{NBDist}
\end{equation}

\noindent where $n = 0, 1, 2,...$, $0 \leq p \leq 1$ and $k$ is a
positive real number. In the application to high energy physics $n$
has the meaning of the number of charged particles detected in an
event. The expected value $\bar{n}$ and variance $V(n)$
\footnote{Here, these quantities are distinguished from the
experimentally measured the average charged particle multiplicity
$\langle N_{ch} \rangle$ and the variance $\sigma^{2}$.
\label{przyp1}} are expressed as:

\begin{equation}
\bar{n} = \frac{k(1-p)}{p}\;,\;\;\;\;\;\;\;\;V(n) =
\frac{k(1-p)}{p^{2}} \;. \label{Parametpk}
\end{equation}

Multiplicity fluctuations are expressed in terms of the scaled
variance:

\begin{equation}
\omega = \frac{\langle N_{ch}^2 \rangle - \langle N_{ch}
\rangle^2}{\langle N_{ch} \rangle} = \frac{V(n)}{\bar{n}}\;,
\label{Chargfluct}
\end{equation}
where $N_{ch}$ is the charged particle multiplicity and the last
equality is valid only for the whole population (the set of all
possible outcomes if the experiment is repeated infinitely many
times), assuming that the hypothesis about the NBD is true.

In application to the high energy physics, the parameters $k,
\bar{n}$ instead of $k, p$ are used usually and

\begin{equation}
\frac{1}{p} = 1 + \frac{\bar{n}}{k} = \omega  \;, \label{Parameterp}
\end{equation}
which is the scaled variance, Eq.~(\ref{Chargfluct}). But because
the centrality bins have the nonzero width, fluctuations defined by
Eq.~(\ref{Chargfluct}) also include a non-dynamical component. This
component is the result of the fluctuations of the geometry of the
collisions within a given centrality bin. The geometrical
fluctuations were evaluated by the PHENIX Collaboration in
\cite{Adare:2008ns}. It turned out that those fluctuations can be
expressed by a correction factor, $f_{geo}$, which is independent of
centrality but varies with the collision type. Then the pure scaled
variance now representing only dynamical fluctuations, i.e. after
subtraction of the geometrical component, can be calculated from the
following equation \cite{Adare:2008ns}:

\begin{equation}
\omega_{dyn} - 1 = f_{geo} \cdot (\omega - 1) \;. \label{Scalvardyn}
\end{equation}
Also parameter $k$ changes to $k_{dyn}$ accordingly

\begin{equation}
k^{-1}_{dyn} = f_{geo} \cdot k^{-1} \;. \label{kadyn}
\end{equation}

In this analysis the hypothesis that the charged-particle
multiplicities measured in ultra-relativistic heavy-ion collisions are
distributed according to the NBD is verified with the use of the
maximum likelihood method (ML) and the likelihood ratio test. More
details of this approach can be found in
Refs.~\cite{Cowan:1998ji,James:2006zz,Baker:1983tu}.

There are two crucial reasons for this approach:

\begin{enumerate}
 \item The fitted quantity is a probability distribution function (p.d.f.),
 so the most natural way is to use the ML method, where the
 likelihood function is constructed directly from the tested p.d.f.. In fact, what is fitted are parameters of the distribution. The fitted values are the \textit{estimators} of these parameters. It is well-known in mathematical statistics that an ML estimator is consistent, asymptotically unbiased and efficient \cite{Cowan:1998ji,James:2006zz,Hoel:1971aa}.
 But even more important is that because of Wilks's theorem (see Appendix~\ref{Wilks}) one can easily define a statistic, the distribution of which
 converges to a $\chi^2$
 distribution as the number of measurements goes to infinity. Thus
 for the large sample the goodness-of-fit can be expressed as a
 $P$-value computed with the corresponding $\chi^2$ distribution.

 \item The most commonly used method, the least-squares (LS) method
 (called also the $\chi^2$ minimization), has the disadvantage of
 providing only the qualitative measure of the significance of the
 fit, in general. Only if observables are represented by Gaussian
 random variables with known variances, the conclusion about
 the goodness-of-fit equivalent to that mentioned in the point 1
 can be derived (see Appendix~\ref{Capsule}).

\end{enumerate}

It is worth noting that the ML method with binned data and Poisson
fluctuations within a bin was already applied to fitting
multiplicity distributions to the NBD but at much lower energies
(E-802 Collaboration \cite{Abbott:1995as}).

\section{Likelihood ratio test}
\label{Liktest}

The number of charged particles $N_{ch}$ is assumed to be a random
variable with the p.d.f. given by Eq.~(\ref{NBDist}). Each event is
treated as an independent observation of $N_{ch}$ and a set of a
given class of events is a sample. For $N$ events in the class there
are $N$ measurements of $N_{ch}$, say $\textbf{X} = \{
X_1,X_2,...,X_N \}$. Some of these measurements can be equal, {\it
i.e.} $X_i = X_j$ for $i \neq j$ can happen. The whole population
consists of all possible events with the measurements of 0, 1, 2,...
charged particles and by definition is infinite \footnote{
Precisely, because of the energy conservation the number of produced
charged particles is limited but the number of collisions is not.
\label{przyp2}}.

Let divide the sample into $m$ bins
characterized by $Y_i$ - the number of measured charged particles
\footnote{Now $Y_i \neq Y_j$ for $i \neq j$ and $i, j = 1, 2,...,m$.
\label{przyp3}} and $n_i$ - the number of entries in the $i$th bin,
$N = \sum_{i=1}^{m}\;n_i$ (details of the theoretical framework of
this Section can be found in
Refs.~\cite{Cowan:1998ji,James:2006zz,Baker:1983tu}). Then the
expectation value of the number of events in the $i$th bin can be
written as

\begin{equation}
\nu_i(\nu_{tot},p,k) = \nu_{tot} \cdot P(Y_i; p, k) \;,
\label{Neventi}
\end{equation}
where $\nu_{tot}$ is the expected number of all events in the
sample, $\nu_{tot} = \sum_{i=1}^{m}\;\nu_i$. This is because one can
treat the number of events in the sample $N$ also as a random
variable with its own distribution - Poisson one. Generally, the
whole histogram can be treated as one measurement of $m$-dimensional
random vector $\textbf{n}=(n_1,...,n_m)$ which has a multinomial
distribution, so the joint p.d.f. for the measurement of $N$ and
$\textbf{n}$ can be converted to the form
\cite{Cowan:1998ji,Baker:1983tu}:

\begin{equation}
f(\textbf{n};\nu_1,...,\nu_m) = \prod_{i=1}^{m}
\frac{\nu_i^{n_i}}{n_i!}\;\exp{(-\nu_i)} \;. \label{jointpdf}
\end{equation}
Since now $f(\textbf{n};\nu_1,...,\nu_m)$ is the p.d.f. for one
measurement, $f$ is also the likelihood function

\begin{equation}
L(\textbf{n} \mid \nu_1,...,\nu_m) = f(\textbf{n};\nu_1,...,\nu_m)
\;. \label{Ljointpdf}
\end{equation}
With the use of Eq.~(\ref{Neventi}) the corresponding likelihood
function can be written as

\begin{equation}
L(\textbf{n} \mid \nu_{tot}, p, k) = L(\textbf{n} \mid
\nu_1(\nu_{tot},p,k),...,\nu_m(\nu_{tot},p,k)) \;. \label{Liksubset}
\end{equation}
Then the likelihood ratio is defined as

\begin{equation}
\lambda = \frac{L(\textbf{n} \mid \hat{\nu}_{tot}, \hat{p},
\hat{k})}{L(\textbf{n} \mid \breve{\nu}_1,...,\breve{\nu}_m)} =
\frac{L(\textbf{n} \mid \hat{\nu}_{tot}, \hat{p},
\hat{k})}{L(\textbf{n} \mid n_1,...,n_m)} \;. \label{Likeliratio}
\end{equation}
where $\hat{\nu}_{tot}$, $\hat{p}$ and $\hat{k}$ are the ML
estimates of $\nu_{tot}$, $p$ and $k$ with the likelihood function
given by Eq.~(\ref{Liksubset}) and $\breve{\nu}_i = n_i$, $i = 1,
2,...m$ are the ML estimates of $\nu_i$ treated as free parameters.
Note that since the denominator in Eq.~(\ref{Likeliratio}) does not
depend on parameters, the log-ratio defined as

\begin{eqnarray}
&&\ln{\lambda(\nu_{tot}, p, k)} = \ln { \frac{L(\textbf{n} \mid
\nu_{tot}, p, k)}{L(\textbf{n} \mid n_1,...,n_m)} } \cr \cr & & = -
\sum_{i=1}^{m}\;\bigg (n_i \ln{\frac{n_i}{\nu_i}} + \nu_i - n_i
\bigg ) \cr \cr & & = - \nu_{tot} + N - \sum_{i=1}^{m}\;n_i
\ln{\frac{n_i}{\nu_i}} \;, \label{Logratio}
\end{eqnarray}
where $\nu_i$ are expressed by Eq.~(\ref{Neventi}), can be used to
find the ML estimates of $\nu_{tot}$, $p$ and $k$. The values $\hat{\nu}_{tot}$, $\hat{p}$ and $\hat{k}$ for which $\lambda(\nu_{tot}, p, k)$ has its maximum are the maximum likelihood estimates of parameters $\nu_{tot}$, $p$ and $k$. Then one can defined the test statistic called ''likelihood $\chi^2$'' \cite{Baker:1983tu}:

\begin{equation}
\chi^2_{\lambda} = -2 \ln{\lambda(\nu_{tot}, p, k)} =
2 \sum_{i=1}^{m}\;\bigg (\nu_i - n_i +
n_i \ln{\frac{n_i}{\nu_i}} \bigg ) \;.
\label{PoissonChi}
\end{equation}
Note that the maximum of $\ln{\lambda}$ is the minimum of
$\chi^2_{\lambda}$, so the estimates from the condition of the minimum of $\chi^2_{\lambda}$ are the ML estimates. Further, the statistic given by

\begin{equation}
\chi^2_{\lambda,min} = -2 \ln{\lambda(\hat{\nu}_{tot}, \hat{p}, \hat{k})} =
2 \sum_{i=1}^{m}\;\bigg (n_i
\ln{\frac{n_i}{\hat{\nu}_i}} + \hat{\nu}_i - n_i \bigg ) \;
\label{PoissonChi}
\end{equation}
approaches a $\chi^2$ distribution asymptotically, i.e. as
the number of measurements, here the number of events $N$, goes to
infinity (the consequence of the Wilks's theorem, see Appendix~\ref{Wilks}).
The values $\hat{\nu}_i$ are the estimates of $\nu_i$ given by

\begin{equation}
\hat{\nu}_i = \hat{\nu}_{tot} \cdot P(Y_i; \hat{p}, \hat{k}) \;
\label{Nevenhati}
\end{equation}
and if one assumes that $\nu_{tot}$ does not depend on $p$ and $k$
then $\hat{\nu}_{tot} = N$. For such a case

\begin{equation}
\sum_{i=1}^{m} \hat{\nu}_i = \sum_{i=1}^{m} n_i \; \label{Nevenexp}
\end{equation}
and Eq.~(\ref{PoissonChi}) becomes

\begin{equation}
\chi^2_{\lambda,min}(\hat{p}, \hat{k}) = 2 \sum_{i=1}^{m}\; n_i
\ln{\frac{n_i}{\hat{\nu}_i}}.  \; \label{MultinomChi}
\end{equation}
%
Also then one can just put $\nu_{tot} = N$ and Eq.~(\ref{Logratio})
can be rewritten as

\begin{eqnarray}
&&\ln{\lambda(p, k)} \cr \cr & & = N \cdot \ln{N} -
\sum_{i=1}^{m}\;n_i \ln{n_i} + \sum_{i=1}^{m}\;n_i \ln{P(Y_i; p, k)}
\cr \cr & & = - \sum_{i=1}^{m}\;n_i \ln{\frac{n_i}{N}} + N
\sum_{i=1}^{m}\;\frac{n_i}{N} \ln{P(Y_i; p, k)} \cr \cr & & = - N
\sum_{i=1}^{m}\;P_i^{ex} \ln{P_i^{ex}} + N \sum_{i=1}^{m}\;P_i^{ex}
\ln{P(Y_i; p, k)}, \label{Logratfreq}
\end{eqnarray}
where $P_i^{ex} = n_i/N$. Thus with the help of
Eqs.~(\ref{MultinomChi}) and (\ref{Logratfreq}) one arrives at

\begin{equation}
\chi^2_{\lambda,min} = 2\; N \sum_{i=1}^{m}\;P_i^{ex} \ln{\frac{P_i^{ex}}{P(Y_i;
\hat{p}, \hat{k})}} \;. \label{FinalChi}
\end{equation}
It can be proven that one of the necessary conditions for the existence of the maximum is (see Appendix~\ref{Aaa} for details):

\begin{equation}
\bar{n} = \langle N_{ch} \rangle \;, \label{eqalaver}
\end{equation}
{\it i.e.} the distribution average has to be equal to the
experimental average. This is very good because $\langle N_{ch} \rangle$ is what is called in statistics \textit{a sample mean}. The sample mean is an estimator for the expectation value of the random variable, which is consistent and unbiased \cite{Cowan:1998ji}. In other words the ML estimator of $\bar{n}$ is $\langle N_{ch} \rangle$ ($\hat{\bar{n}} = \langle N_{ch} \rangle$).

\section{Results and discussion}
\label{Finl}

The method described in Sec.~\ref{Liktest}
requires that all bins in a given data set have the width equal to
1, so as the experimental probability $P_i^{ex}$ to measure a signal
in the $i$th bin was equivalent to the probability of the
measurement of $(i-1)$ charged particles (the first bin is the bin
of 0 charged particles detected). This is fulfilled for all bins of
the considered data sets

Since the test statistic $\chi^2_{\lambda,min}$ has a $\chi^{2}$
distribution approximately in the large sample limit, it can be used
as a test of the goodness-of-fit. The result of the test is given by
the so-called $P$-value which is the probability of obtaining the
value of the statistic, Eq.~(\ref{PoissonChi}), equal to or greater
then the value just obtained for the present data
set, when repeating the whole experiment many times (see Appendix~\ref{Capsule}):

\begin{equation}
P = P(\chi^{2} \geq  \chi^2_{\lambda,min}; n_d) =
\int_{\chi^2_{\lambda,min}}^{\infty}\;f(z;n_d) dz \;, \label{Pvalue}
\end{equation}
where $f(z;n_d)$ is the $\chi^{2}$ p.d.f. and $n_d$ the number of
degrees of freedom, $n_d = m-2$ here.

\begin{figure}
 \begin{center}
  \begin{tabular}{c c}
  {\resizebox{!}{8cm}{\includegraphics{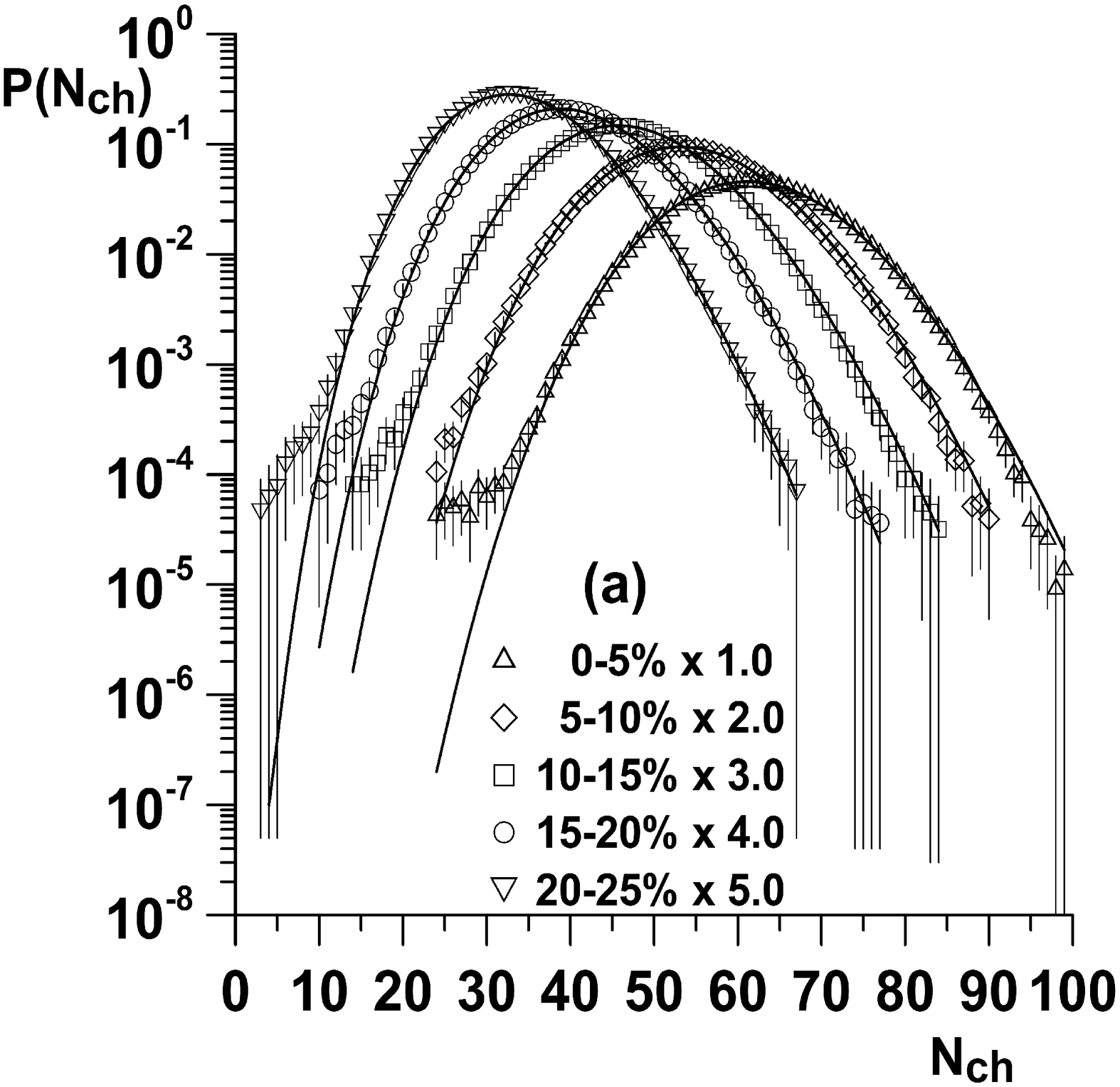}}} & $\;\;\;\;\;\;$
  {\resizebox{!}{8cm}{\includegraphics{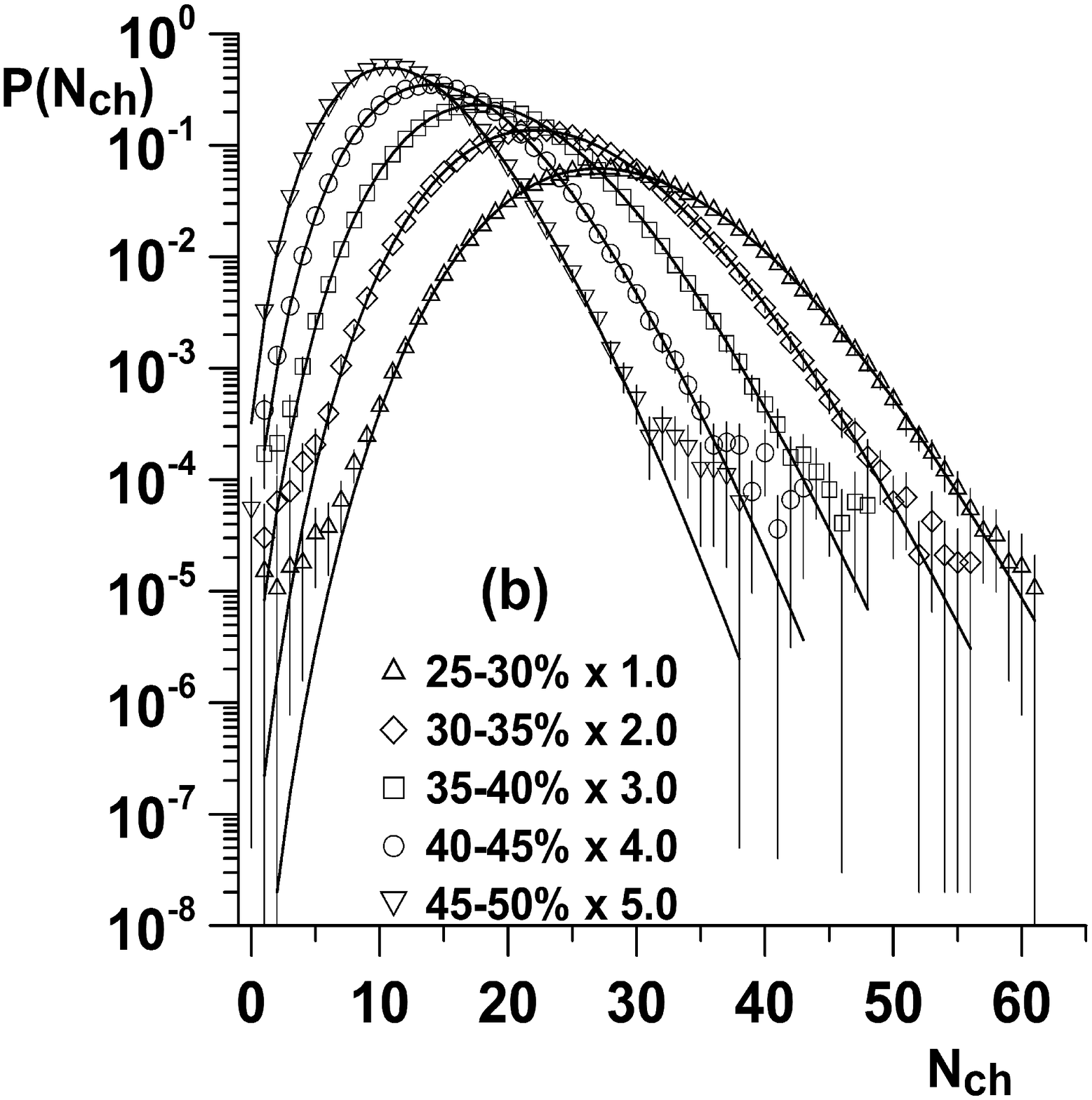}}}
  \end{tabular}
    \caption{ Uncorrected multiplicity distributions of charged hadrons
     for 200 GeV Au-Au collisions \protect\cite{Adare:2008ns} within ranges
     limited to the bins with $n_{i} > 5$. The lines are fits to the NBD.
     The data are scaled by the amounts in the legend. Errors represent
     the statistical and systematic errors added in quadrature. }
  \label{fig1}
 \end{center}
\end{figure}

\begin{figure}
 \begin{center}
  \begin{tabular}{c c}
  {\resizebox{!}{8cm}{\includegraphics{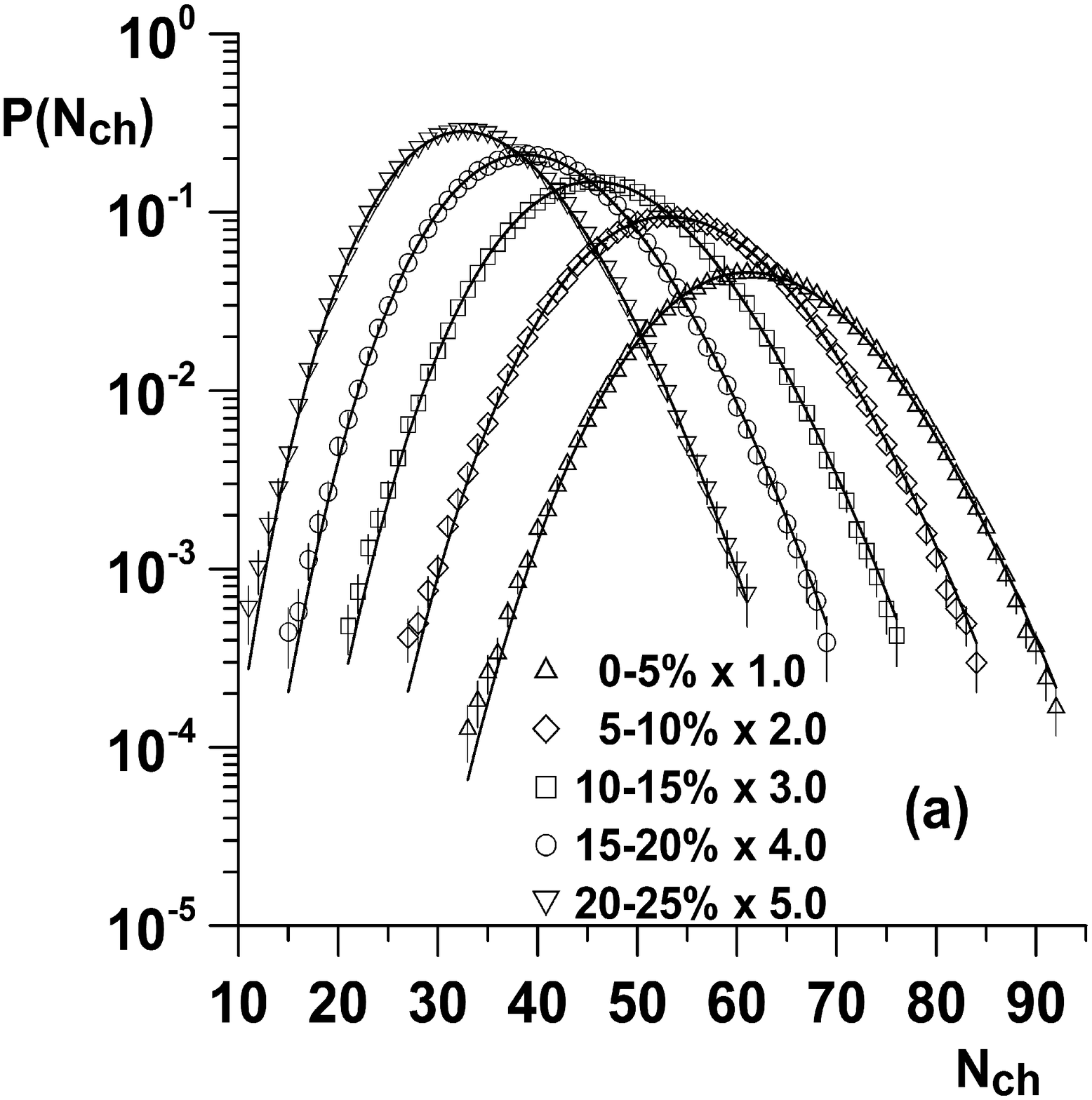}}} & $\;\;\;\;\;\;$
  {\resizebox{!}{8cm}{\includegraphics{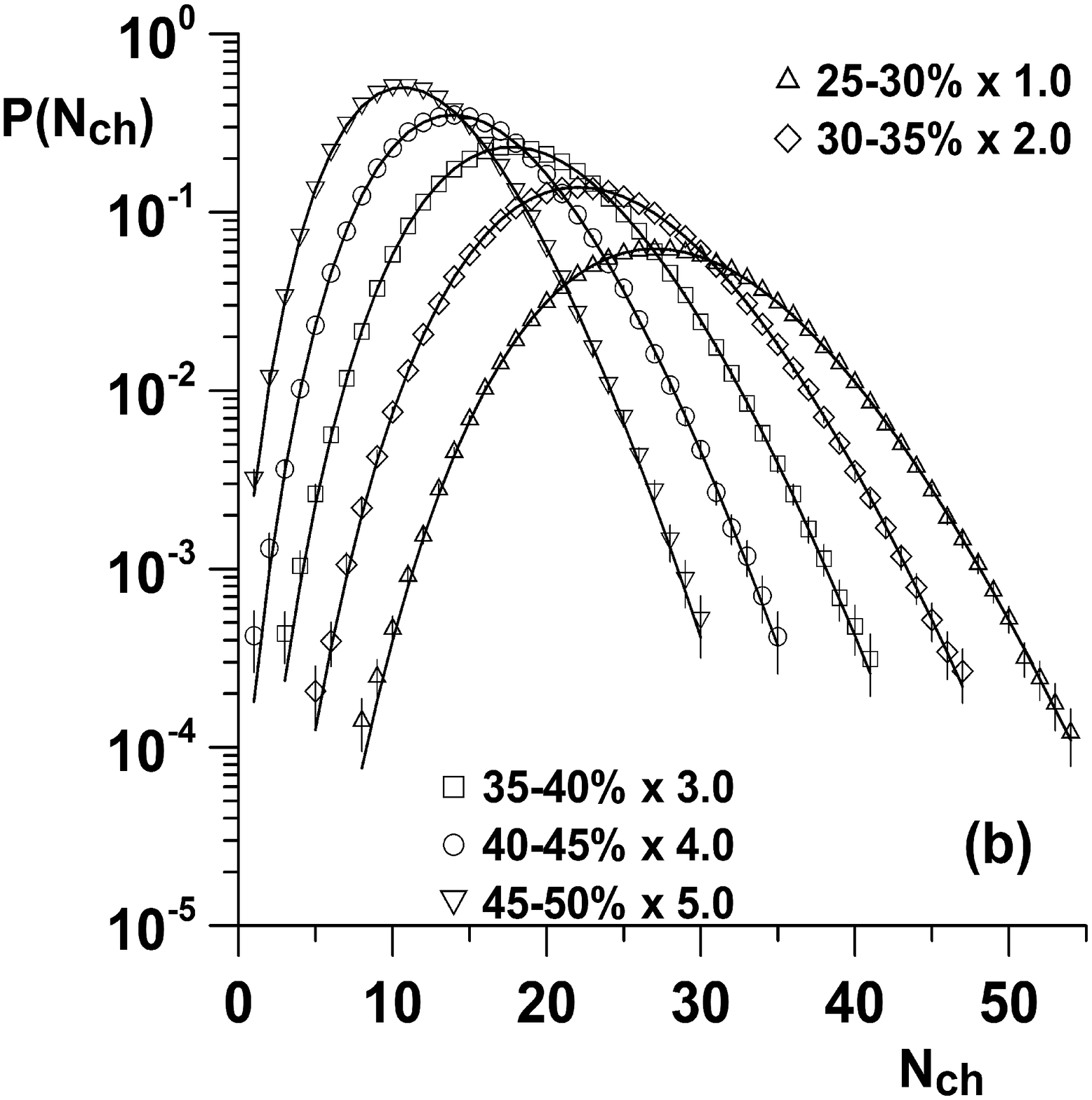}}}
  \end{tabular}
    \caption{ Uncorrected multiplicity distributions of charged hadrons
     for 200 GeV Au-Au collisions \protect\cite{Adare:2008ns} within ranges
     limited to the bins with $n_{i} > 60$. The lines are fits to the NBD.
     The data are scaled by the amounts in the legend. Errors represent
     the statistical and systematic errors added in quadrature. }
  \label{fig2}
 \end{center}
\end{figure}

\begin{figure}
 \begin{center}
  \begin{tabular}{c c}
  {\resizebox{!}{8cm}{\includegraphics{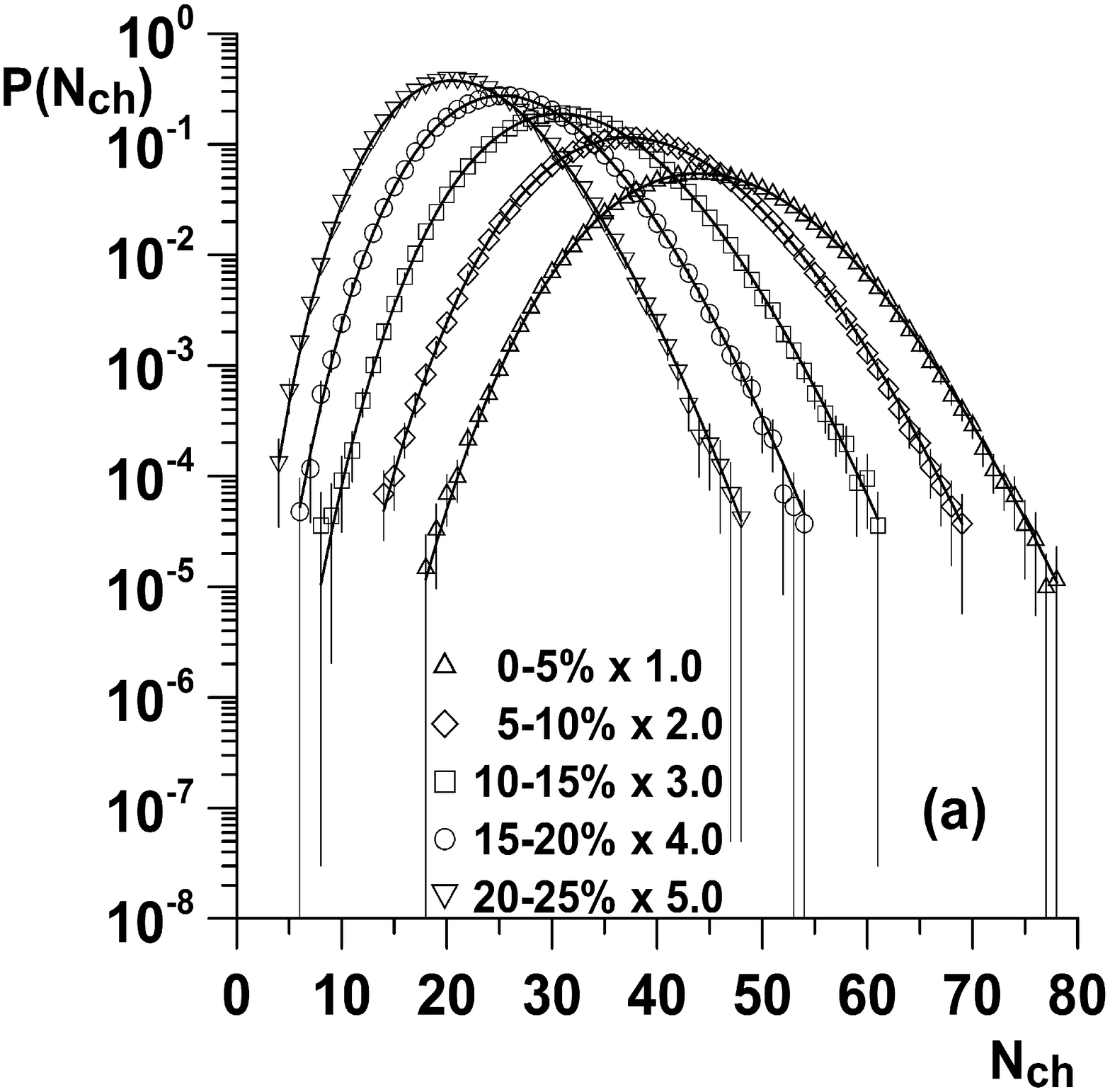}}} & $\;\;\;\;\;\;$
  {\resizebox{!}{8cm}{\includegraphics{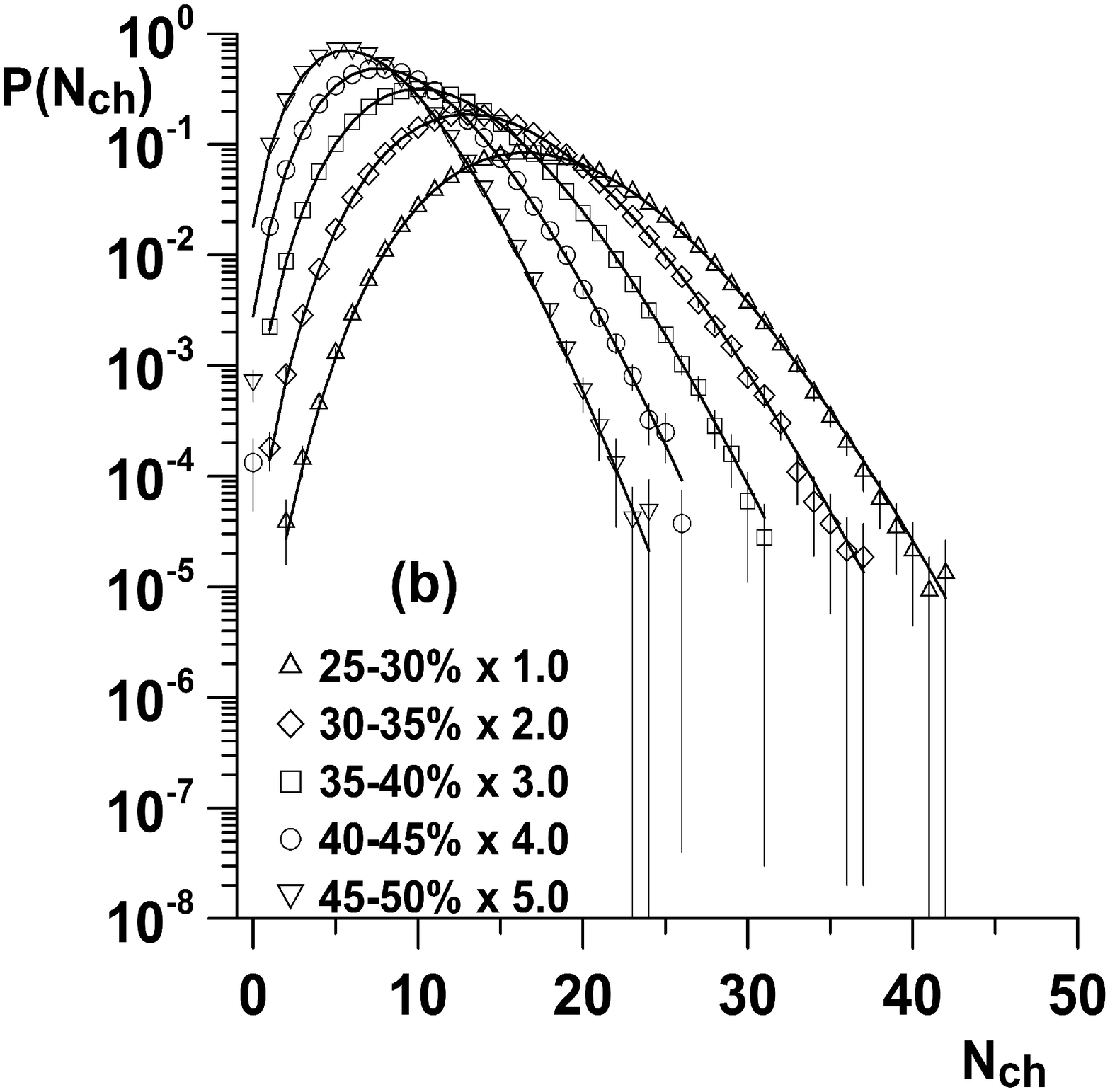}}}
  \end{tabular}
    \caption{ Uncorrected multiplicity distributions of charged hadrons
     for 62.4 GeV Au-Au collisions \protect\cite{Adare:2008ns} within ranges
     limited to the bins with $n_{i} > 5$. The lines are fits to the NBD.
     The data are scaled by the amounts in the legend. Errors represent
     the statistical and systematic errors added in quadrature. }
  \label{fig3}
 \end{center}
\end{figure}

\begin{figure}
 \begin{center}
  \begin{tabular}{c c}
  {\resizebox{!}{8cm}{\includegraphics{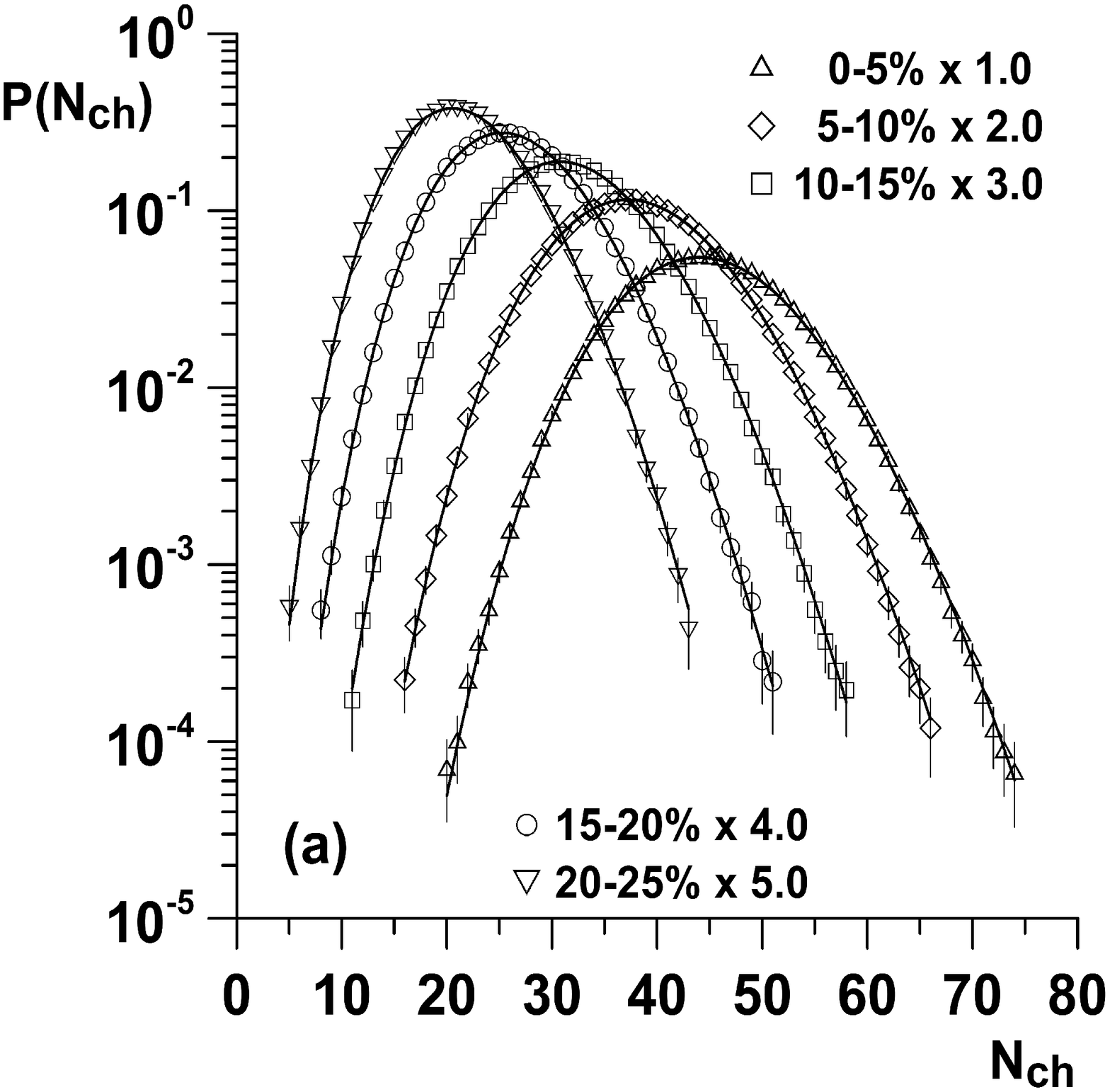}}} & $\;\;\;\;\;\;$
  {\resizebox{!}{8cm}{\includegraphics{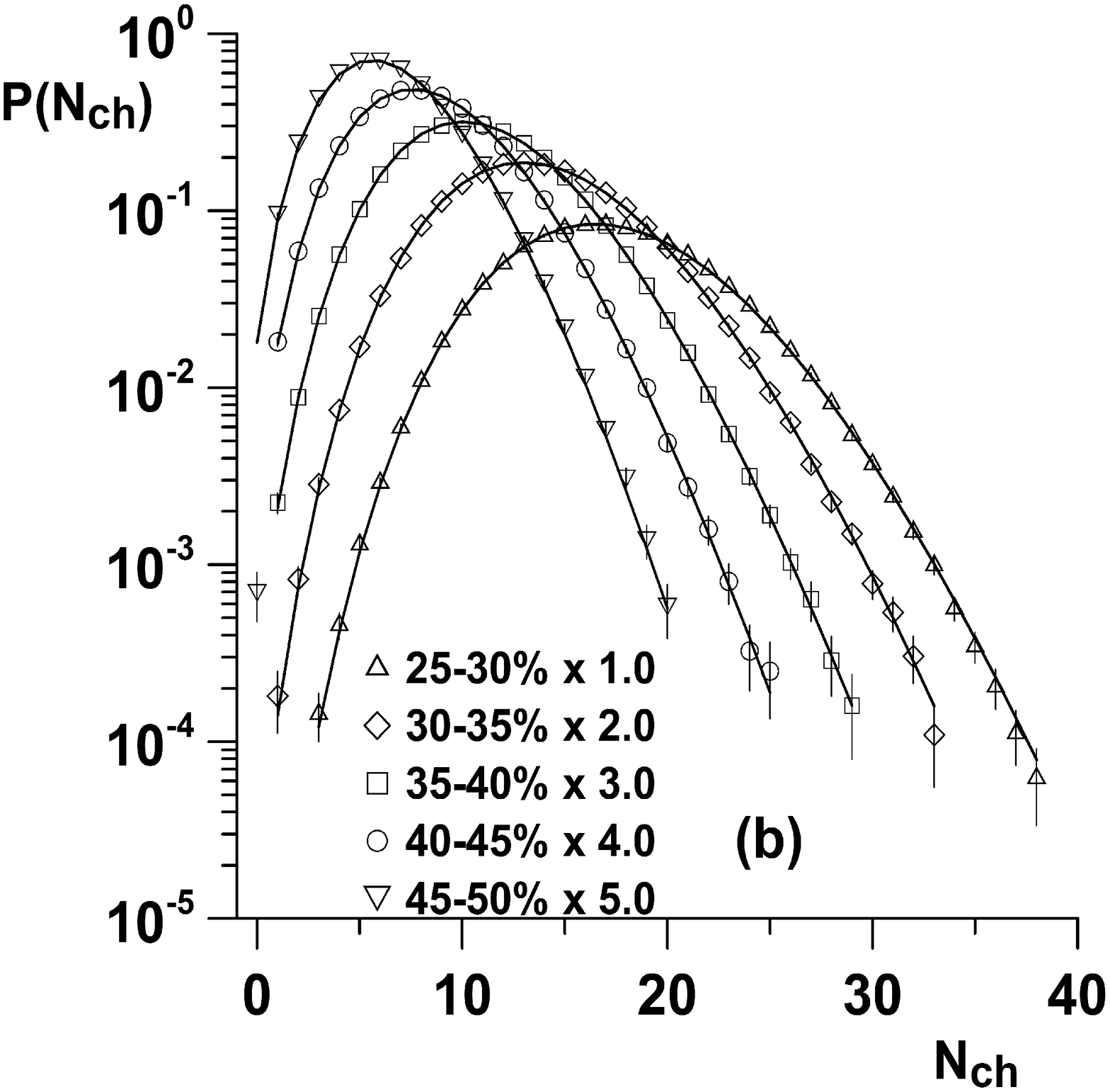}}}
  \end{tabular}
    \caption{ Uncorrected multiplicity distributions of charged hadrons
     for 62.4 GeV Au-Au collisions \protect\cite{Adare:2008ns} within ranges
     limited to the bins with $n_{i} > 40$. The lines are fits to the NBD.
     The data are scaled by the amounts in the legend. Errors represent
     the statistical and systematic errors added in quadrature. }
  \label{fig4}
 \end{center}
\end{figure}

\begin{figure}
 \begin{center}
  \begin{tabular}{c c}
  {\resizebox{!}{8cm}{\includegraphics{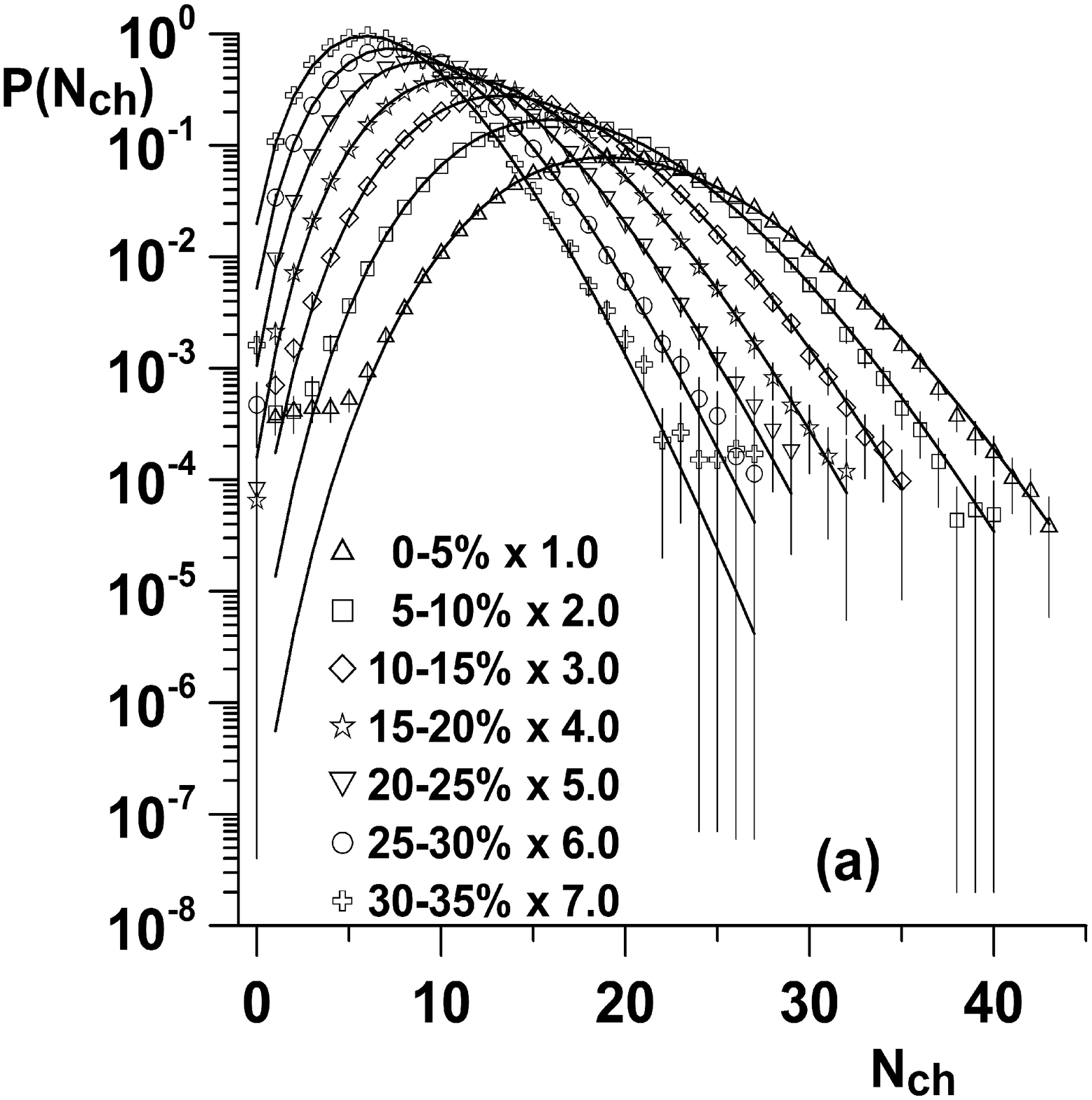}}} & $\;\;\;\;\;\;$
  {\resizebox{!}{8cm}{\includegraphics{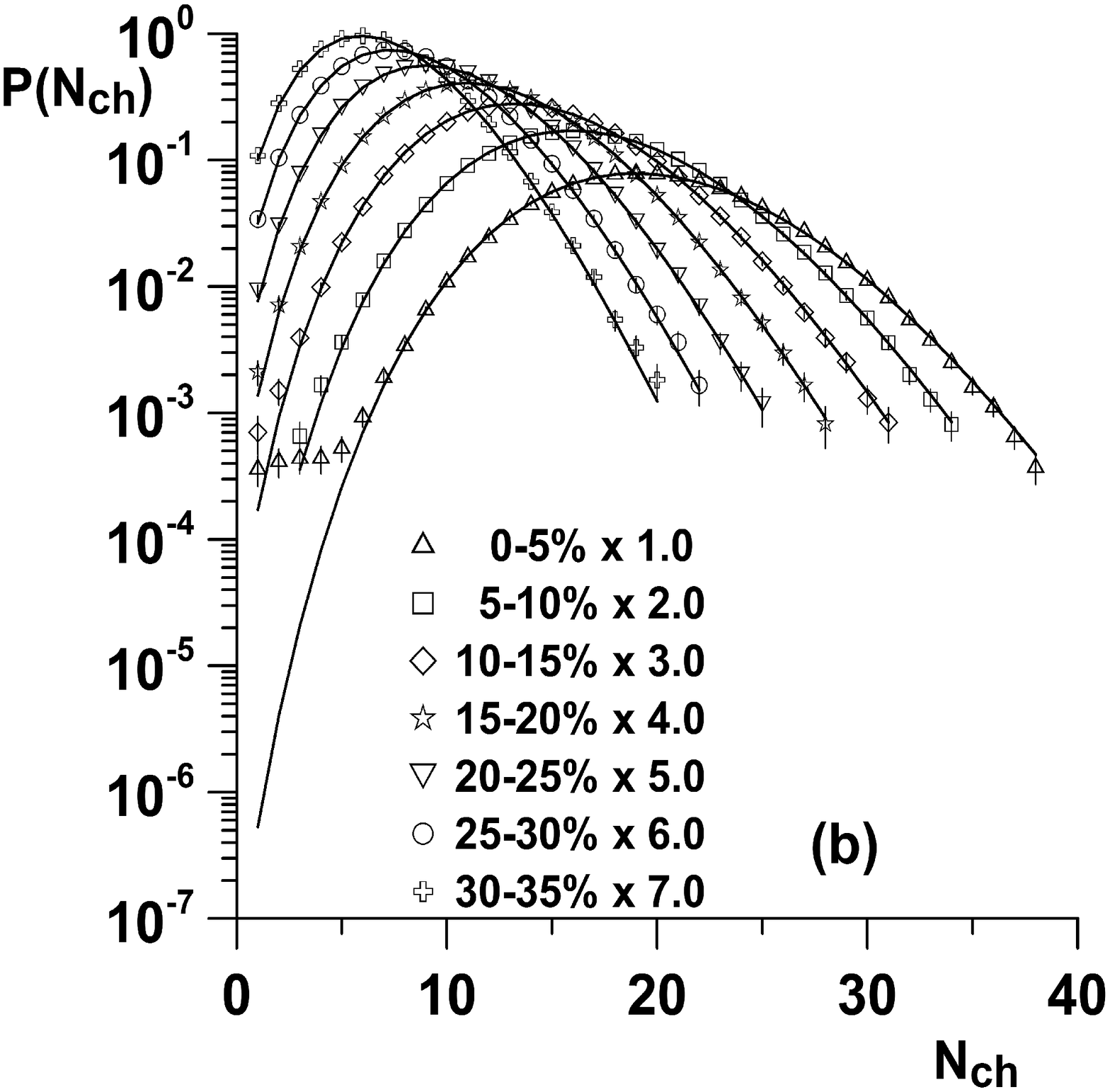}}}
  \end{tabular}
    \caption{ Uncorrected multiplicity distributions of charged hadrons
     for 200 GeV Cu-Cu collisions \protect\cite{Adare:2008ns} within ranges
     limited to the bins with $n_{i} > 5$ (left) and $n_{i} > 80$ (right).
     The lines are fits to the NBD.
     The data are scaled by the amounts in the legend. Errors represent
     the statistical and systematic errors added in quadrature. }
  \label{fig5}
 \end{center}
\end{figure}

\begin{figure}
 \begin{center}
  \begin{tabular}{c c}
  {\resizebox{!}{8cm}{\includegraphics{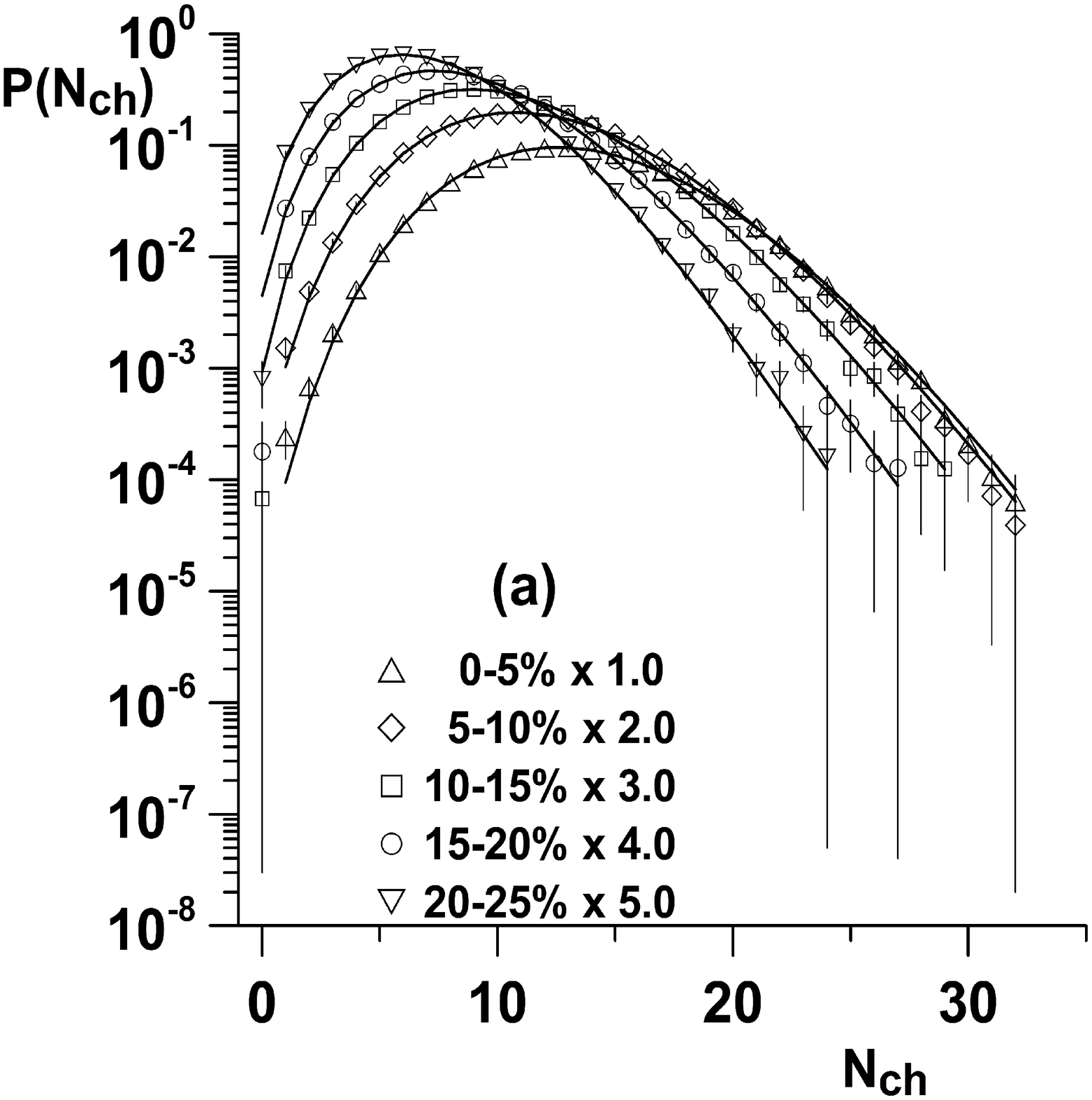}}} & $\;\;\;\;\;\;$
  {\resizebox{!}{8cm}{\includegraphics{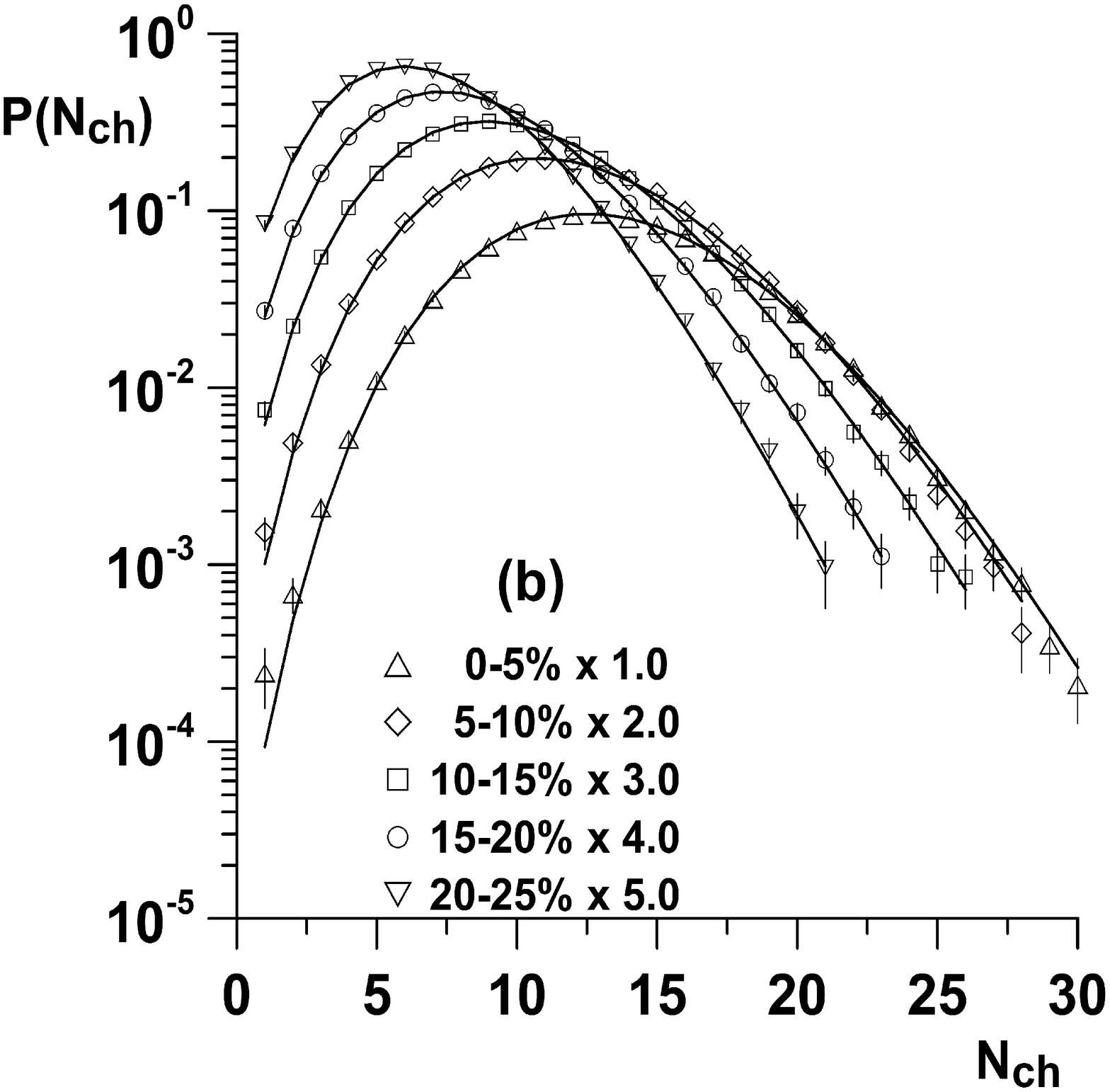}}}
  \end{tabular}
    \caption{ Uncorrected multiplicity distributions of charged hadrons
     for 62.4 GeV Cu-Cu collisions \protect\cite{Adare:2008ns} within ranges
     limited to the bins with $n_{i} > 5$ (left) and $n_{i} > 60$ (right).
     The lines are fits to the NBD.
     The data are scaled by the amounts in the legend. Errors represent
     the statistical and systematic errors added in quadrature. }
  \label{fig6}
 \end{center}
\end{figure}

The results of the analysis are presented in Tables~\ref{Table1}-\ref{Table8} and illustrated with Figs.~\ref{fig1}-\ref{fig6}. In fact the whole analysis was done for the two kinds of histograms: (\textit{i}) bins with the number of entries $n_{i} \leq 5$ excluded, Tables~\ref{Table1}, \ref{Table3}, \ref{Table5} and \ref{Table7}; (\textit{ii}) bins with the number of entries $n_{i} \leq 40$, Table~\ref{Table4}, $n_{i} \leq 60$, Tables~\ref{Table2} and \ref{Table8} or $n_{i} \leq 80$, Table~\ref{Table6}, excluded. In practice this corresponds to cutting off less (\textit{i}) or more (\textit{ii}) the tails of the full measured histograms. The tails break the visual agreement between the data and the NBD, cf. Figs.~\ref{fig1} and \ref{fig2}. The condition that only bins with $n_{i} > 5$ are taken into account is the minimal condition imposed on a histogram to do any statistical inference without Monte Carlo simulations \cite{Cowan:1998ji}. The condition (\textit{ii}) corresponds roughly to the choice made originally by the PHENIX Collaboration in their analysis \cite{Adare:2008ns}. It has turned out that the results of this analysis are qualitatively the same for both choices.

As one can see, the hypothesis in question should
be rejected in all considered cases. But it was claimed that charged-particle multiplicities measured in Au-Au and Cu-Cu collisions at $\sqrt{s_{NN}} = 62.4$ and
200 GeV are distributed according to the NBD \cite{Adare:2008ns}.
However that conclusion was the result of the application of
the LS method. Therefore it seems to be reasonable to check what
are the values of the LS test statistic at the ML estimators
listed in the third and fourth columns of Tables~\ref{Table1}-\ref{Table8}. For
the sample described in Sec.~\ref{Liktest} one can define the LS test statistic
(commonly called the $\chi^{2}$ function) as:

\begin{equation}
\chi_{LS}^{2}(\textbf{n};\bar{n},k) = \sum_{i=1}^{m} \frac{(n_i-\nu_i(\bar{n}, k)
)^{2}}{err_{n_i}^{2}} = \sum_{i=1}^{m} \frac{(P_i^{ex}-P(Y_i; \bar{n}, k)
)^{2}}{err_{i}^{2}}   \;, \label{Chidef}
\end{equation}

\noindent where $\nu_i(\bar{n}, k)$ is given by Eq.~(\ref{Neventi}) with $\nu_{tot} = N$ and $err_{n_i}$ ($err_{i} = err_{n_i}/N$) is the uncertainty on $n_i$ ($P_i^{ex}$ respectively). Note that for $err_{n_i}^{2} = \nu_i$ the above equation is the Pearson's $\chi^2$ test statistic, whereas for $err_{n_i}^{2} = n_i$ this is the Neyman's $\chi^2$ test statistic (also called the modified chi-square or modified least-squares method), both well known in mathematical statistics \cite{Cowan:1998ji,James:2006zz,Baker:1983tu,Kendall:1999bb}. The advantage of the use of these statistics is that both follow a $\chi^2$ distribution asymptotically. The errors given by $\sqrt{\nu_i}$ or $\sqrt{n_i}$ are interpreted as theoretical or experimental statistical errors respectively (for the discussion of the pros and cons of both see \cite{Cowan:1998ji,Lyons:1986en}). It should be stressed that when $err_{n_i}$ includes also a systematic error (e.g. by adding in quadrature to statistical one), then the statement about asymptotic form of the distribution of the test statistic is no longer valid.

In the present analysis $\chi_{LS}^{2}$ function, Eq.~(\ref{Chidef}), \textbf{is not minimized} with respect to $\bar{n}$ and $k$ (or $p$ and $k$) as in the LS method but is calculated at ML estimates of $\bar{n}$ and $k$. Generally, this is allowed in statistics and is equivalent to test a single point in the parameter space. Then
the tested point might not be the best estimate of the true value
but the hypothesis in question becomes the hypothesis only about a
particular distribution (a \textit{simple} hypothesis).
At first sight, $\chi_{LS}^2$/$n_d$ values of the ninth column of Tables~\ref{Table1}-\ref{Table8} seem to be significant for almost all
centrality classes, what agrees with the results of
Ref.~\cite{Adare:2008ns}. But this contradicts the results of the likelihood ratio test, which are expressed by $\chi^2_{\lambda}$/$n_d$ and $P$-values listed in the seventh and eight columns of Tables~\ref{Table1}-\ref{Table8}. The crucial question is now why the conclusions from $\chi^2_{\lambda}$ and $\chi_{LS}^2$ test statistics are entirely opposite for PHENIX measurements?
The main difference between both statistics is that $\chi^2_{\lambda}$ does not depend on the actual errors but $\chi_{LS}^2$ does. Additionally,
$\chi^2_{\lambda}$ depends explicitly on the number of events whereas $\chi_{LS}^2$ does not, cf. Eqs.~(\ref{FinalChi}) and (\ref{Chidef}). In principle, one can conclude that $\chi^2_{\lambda}$ statistic implicitly assumes errors of the type $\sqrt{n_i}$ because the statistic originated from the likelihood function,
Eqs.~(\ref{jointpdf}) and (\ref{Ljointpdf}), which is the product of
Poisson distributions. But there is no place to insert actual experimental errors into $\chi^2_{\lambda}$ statistic, Eqs.~(\ref{Logratio}) and (\ref{PoissonChi}), this test statistic does not take by definition the experimental errors into account. And last but not least, the distribution of $\chi^2_{\lambda,min}$ is known asymptotically, whereas the distribution of $\chi_{LS}^2$ at the minimum, when systematic errors are included, is not known, even asymptotically.

In the PHENIX analysis \cite{Adare:2008ns} errors $err_{i}$ in Eq.~(\ref{Chidef}) are represented by the quadrature sum of the statistical and systematic components, the statistical error on the number of entries $n_i$ is equal to $\sqrt{n_i}$ exactly \cite{ppg070:data} (the statistical error on $P_i^{ex}$ is $\sqrt{n_i}/N$ then). The systematic errors were mostly caused by time-dependent variation of results. Data sets were taken over spans of several days to several weeks, during which the total acceptance and efficiency were changing, mainly because of degradation of the tracking detectors \cite{Adare:2008ns,Mitchell:2011pp}. To estimate these systematic errors, the entire data set was divided into 10 subsets of approximately equal sizes. Then plots from these subsets were overlaid with each other, from which bin-by-bin systematic errors were estimated as 3.0 times the statistical errors, the same for all data sets and centralities \cite{ppg070:data,Mitchell:2011pp} \footnote{This detailed  information is from Ref.~\cite{Mitchell:2011pp}, but there is a short note:  {\it''On average, the systematic + statistical errors are a factor of
3 larger than the statistical errors.''} in Ref.~\cite{ppg070:data}.
\label{przyp4}}. This causes that $err_{n_i}^2 = \sigma_{stat,n_i}^2 + 9 \cdot \sigma_{stat,n_i}^2 = 10 \cdot \sigma_{stat,n_i}^2 = 10 \cdot n_i$ ($err_{n_i} = \sqrt{10} \cdot \sigma_{stat,n_i} \approx 3.0 \cdot \sigma_{stat,n_i}$), where $\sigma_{stat,n_i} = \sqrt{n_i}$ is the statistical error of the \textit{i}th measurement. Hence if statistical errors only were taken into account the values of $\chi_{LS}^2$/$n_d$ would be 10 times greater than those listed in Tables~\ref{Table1}-\ref{Table8}. So it seems that the acceptance of the NBD hypothesis by $\chi_{LS}^2$ test is entirely due to the magnitude of systematic errors. But in fact this is the result of confused inference as it will be shown further.

If one inserts explicit values of PHENIX errors, $err_{n_i}^2 = 10 \cdot n_i$, into Eq.~(\ref{Chidef}), then $\chi_{LS}^2$ test statistic takes the form called $\chi_{PHEN}^{2}$ from now on (the author strongly advices to read Appendix~\ref{Capsule} first, before going further):

\begin{equation}
\chi_{PHEN}^{2}(\textbf{n};\bar{n},k) = \sum_{i=1}^{m} \frac{(n_i-\nu_i(\bar{n}, k)
)^{2}}{10 \cdot n_i} = \frac{1}{10} \cdot \sum_{i=1}^{m} \frac{(n_i-\nu_i(\bar{n}, k)
)^{2}}{n_i} = \frac{1}{10} \cdot \chi_N^{2}(\textbf{n};\bar{n},k) \;. \label{ChiPHEN}
\end{equation}
But this exactly is the Neyman's $\chi^2$ test statistic, $\chi_N^{2}$, multiplied by 0.1. Therefore PHENIX test statistic estimators of parameters $\bar{n}$ and $k$ are Neyman's $\chi^2$ estimators, $\hat{\bar{n}}_N$ and $\hat{k}_N$ respectively. Further, the distribution of the Neyman's $\chi^2$ test statistic $t_N(\textbf{n}) \equiv \chi_N^{2}(\textbf{n};\hat{\bar{n}}_N,\hat{k}_N)$ asymptotically approaches a $\chi^2$ distribution with $n_d = m-2$ \cite{Baker:1983tu,Berkson:1980chi,Beaujean:2011phst}. Now, the more rigorous justification for inserting ML estimates into $\chi_{LS}^{2}$, Eq.~(\ref{Chidef}), can be given. The likelihood $\chi^2$, Pearson's $\chi^2$ and Neyman's $\chi^2$ test statistics are asymptotically equivalent, i.e. their estimators are consistent, asymptotically normal, with the same minimum variance (Ref.~\cite{James:2006zz}, p. 192; Ref.~\cite{Kendall:1999bb}, Sec. 18.58; Ref.~\cite{Berkson:1980chi}, pp. 457-458). Moreover, \textit{''So far as the $\chi^2$'s considered for tests of significance are concerned, any can be used with any of the estimates''} (Ref.~\cite{Berkson:1972inf}, p. 464; also see p. 444). This means that e.g. ML estimates could be put into the Neyman's $\chi^2$ test statistic and still the distribution of such test statistic would approach a $\chi^2$ distribution asymptotically. Since PHENIX samples are very large (see the second column in Tables~\ref{Table1}-\ref{Table8}) one can reasonably approximate the distribution of $t_N(\textbf{n})$ by the corresponding $\chi^2$ distribution. But what is the distribution of the PHENIX test statistic $t_{PHEN}(\textbf{n}) = \chi_{PHEN,min}^{2} (\textbf{n}) \equiv \chi_{PHEN}^{2}(\textbf{n};\hat{\bar{n}}_N,\hat{k}_N) = 0.1 \cdot t_N(\textbf{n})$ then ? This can be easily done with the use of the general rule of finding the distribution $g(t)$ of a function $t(z)$ of a random variable $z$ with the known p.d.f. $f(z)$ (Ref.~\cite{Cowan:1998ji}, p. 14):

\begin{equation}
g(t) = f(z(t)) \left| \frac{dz}{dt}  \right| \;, \label{distroffunct}
\end{equation}
%
if $t(z)$ has a unique inverse. In the present case $t(z) = 0.1 z$ and $f(z) = f(z;n_d)$, so $z(t) = 10t$ and $g(t;n_d) = 10f(10t;n_d)$ is the distribution in question. The expectation value of the PHENIX test statistic is $E[t_{PHEN}] = E[0.1 \cdot t_N] = 0.1 \cdot E[t_N] = 0.1 \cdot n_d$. Thus $E[t_{PHEN}/n_d] = 0.1$ or rewriting it a in more familiar way: $E[\chi_{PHEN}^{2}/n_d] = 0.1$, \textbf{NOT 1}. Therefore, if the (PHENIX) experiment is 'reasonable' and the hypothesis is true, one should expect to obtain $\chi_{PHEN}^{2}/n_d \approx 0.1$ - values of $\chi_{PHEN}^{2}/n_d$ much greater than 0.1 suggests that the hypothesis (of the NBD) should be rejected. In the language of Appendix~\ref{Capsule}, the decision boundary for the PHENIX test statistic $\chi_{PHEN}^{2}$ should be placed at $0.1 \cdot n_d$, \textbf{NOT at $\textbf{n}_\textbf{d}$}. In the case of $\chi_{PHEN}^{2}$ statistic the $P$-value for the hypothesis is given by

\begin{equation}
P = \int_{t_{PHEN}}^{\infty} g(t;n_d)dt = \int_{10\cdot\chi_{PHEN,min}^{2}}^{\infty} f(t;n_d)dt\;,
\label{PHENPvalue}
\end{equation}
%
where $f(z;n_d)$ is the $\chi^{2}$ p.d.f. with $n_d$ degrees of freedom. The corresponding values are given in the tenth column of Tables~\ref{Table1}-\ref{Table8}.
Altogether there are 33 classes of collision systems and centralities of the PHENIX measurements \cite{Adare:2008ns} considered here. They are doubled because of two possibilities of cutting tails in full histograms. The assessment of the quality of fits presented in Tables~\ref{Table1}-\ref{Table8} depends on the assumed significance level. Following the choice done by the UA5 Collaboration \cite{Ansorge:1988kn}, the $0.1 \%$ level is fixed here. There are 8 cases where the PHENIX test is significant at the $0.1 \%$ level at least for one of the two histograms corresponding to the same class. It is interesting that half of them belong to the case of Au-Au collisions at $\sqrt{s_{NN}} = 62.4$ GeV and are significant for both kinds of histograms with $P$-values greater than $1 \%$, see Tables~\ref{Table3} and \ref{Table4}. The next two happen for Au-Au collisions at $\sqrt{s_{NN}} = 200$ GeV, Table~\ref{Table2}, and the last two for Cu-Cu collisions at $\sqrt{s_{NN}} = 200$ GeV, Table~\ref{Table6}, but only in the case of narrower histograms and with $P$-values smaller than $1 \%$. On opposite, the case of Cu-Cu collisions at $\sqrt{s_{NN}} = 62.4$ GeV has no any significant fit at all, see Tables~\ref{Table7} and \ref{Table8}. Thus one can conclude that only for the PHENIX collision system Au-Au at $\sqrt{s_{NN}} = 62.4$ GeV the hypothesis of the NBD could not be rejected. For other systems the hypothesis of the NBD seems to be very unlikely. What distinguishes the case of Au-Au collisions at $\sqrt{s_{NN}} = 62.4$ GeV from others? The only thing which can be noticed from Tables~\ref{Table1}-\ref{Table8} is that the number of events is substantially greater (about $14 \%$) in this case.

In principle, the accuracy with which experimental distributions approximate the NBD should increase with the sample size because if the hypothesis is true the postulated form of distribution is exact for the whole population. So with the growing
number of events, the experimental distribution should be closer to
the postulated one. This is also seen in the form of $\chi^2_{\lambda,min}$,
Eq.~(\ref{FinalChi}), where the linear dependence on $N$ is
explicit. To keep $\chi^2_{\lambda,min}$ at least constant when $N$ (the
sample size) is growing the relative differences between $P(Y_i)$
and $P_i^{ex}$ have to decrease. The PHENIX test statistic $\chi_{PHEN}^{2}$, Eq.~(\ref{ChiPHEN}), reveals the same feature because relative errors behave like $\sqrt{n_i}/N$. So the results of fits for the collision system Au-Au at $\sqrt{s_{NN}} = 62.4$ GeV are even more valuable.

\begin{figure}
 \begin{center}
  \begin{tabular}{c c}
  {\resizebox{!}{8cm}{\includegraphics{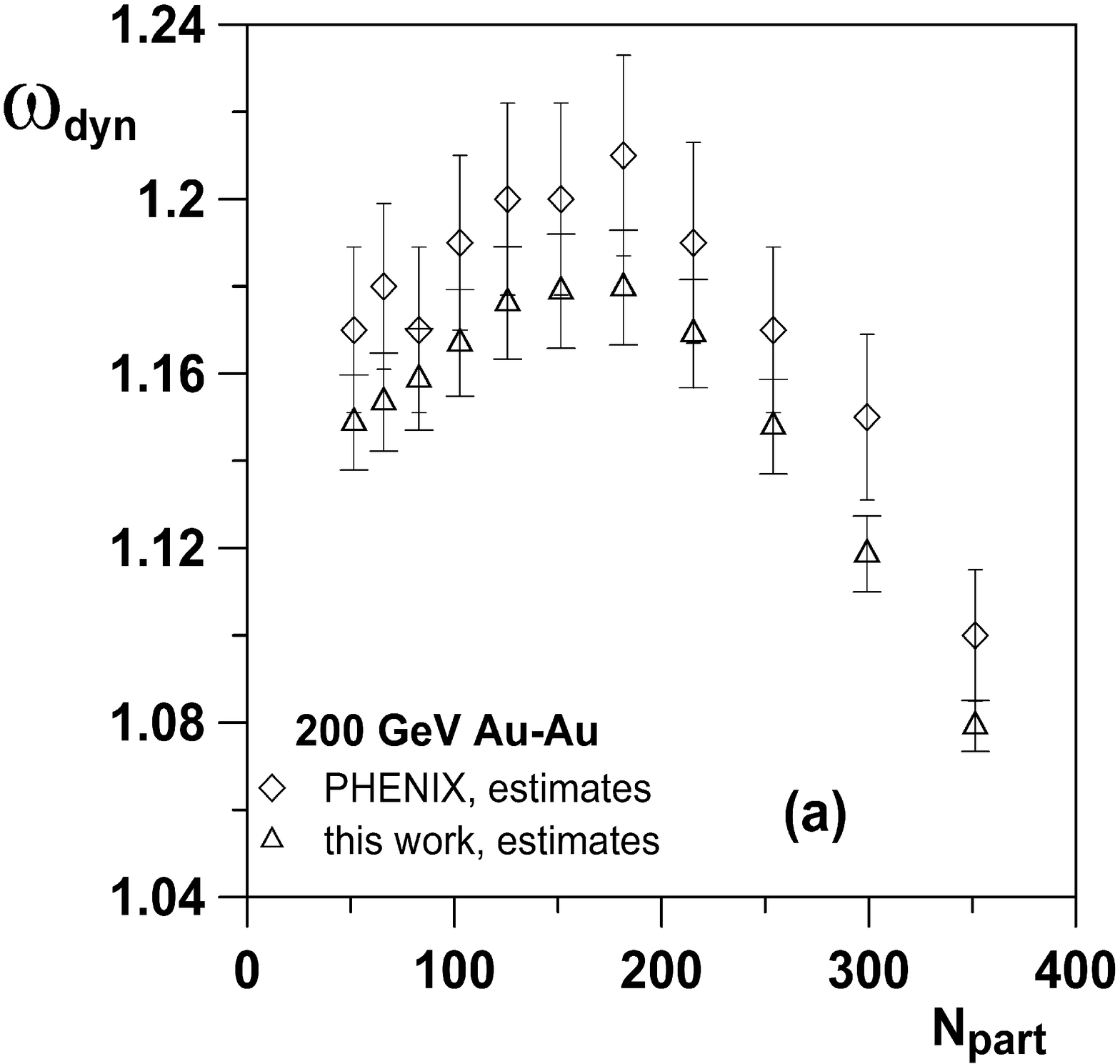}}} & $\;\;\;\;\;\;$
  {\resizebox{!}{8cm}{\includegraphics{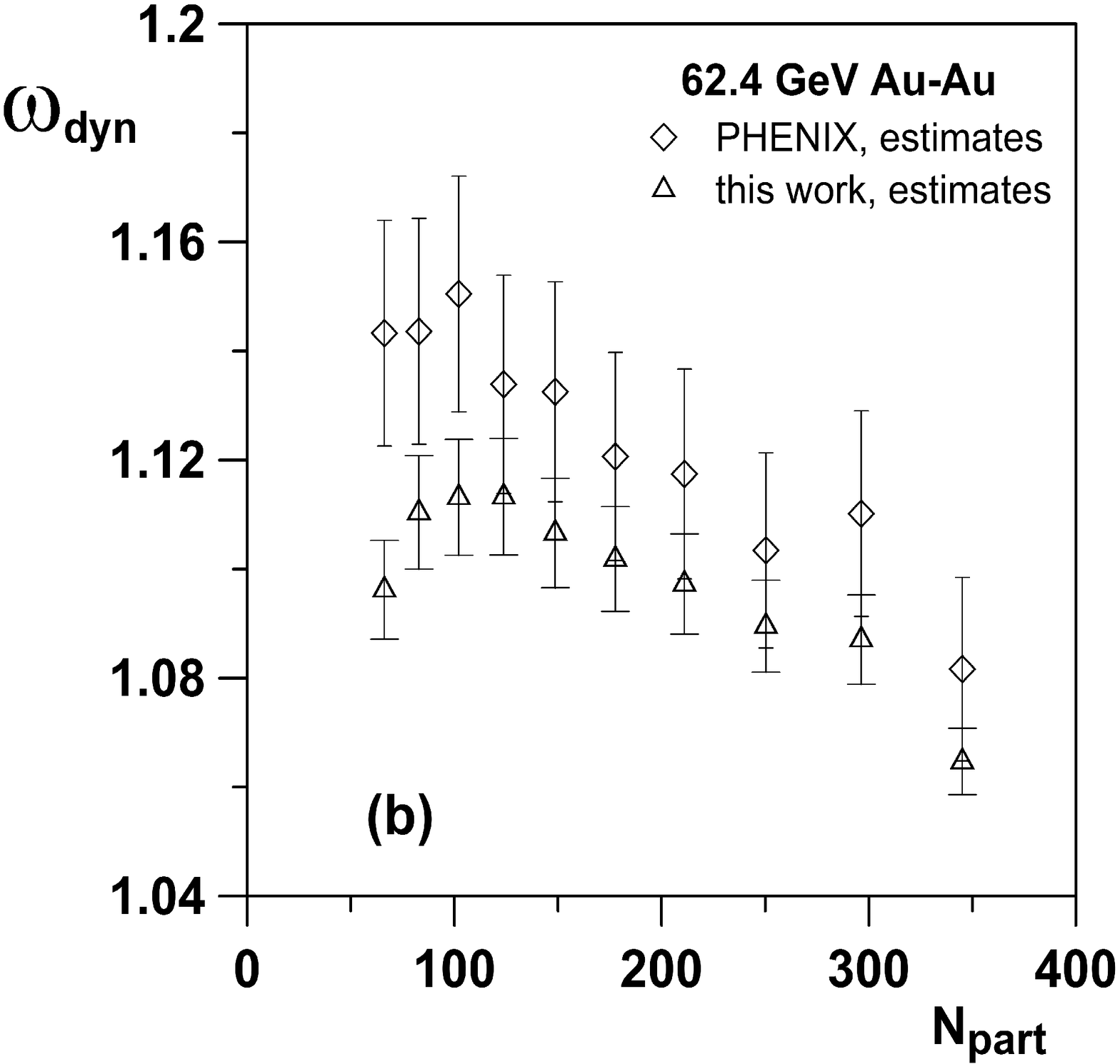}}}
  \end{tabular}
    \caption{ Scaled variance for 200 GeV (a) and 62.4 GeV (b) Au-Au collisions. PHENIX estimates are from \protect\cite{Adare:2008ns}. Estimates from this work are for the cases with ranges limited to the bins where $n_{i} > 60$, see Tables~\ref{Table2} and \ref{Table4}.}
  \label{fig7}
 \end{center}
\end{figure}

\begin{figure}
 \begin{center}
  \begin{tabular}{c c}
  {\resizebox{!}{8cm}{\includegraphics{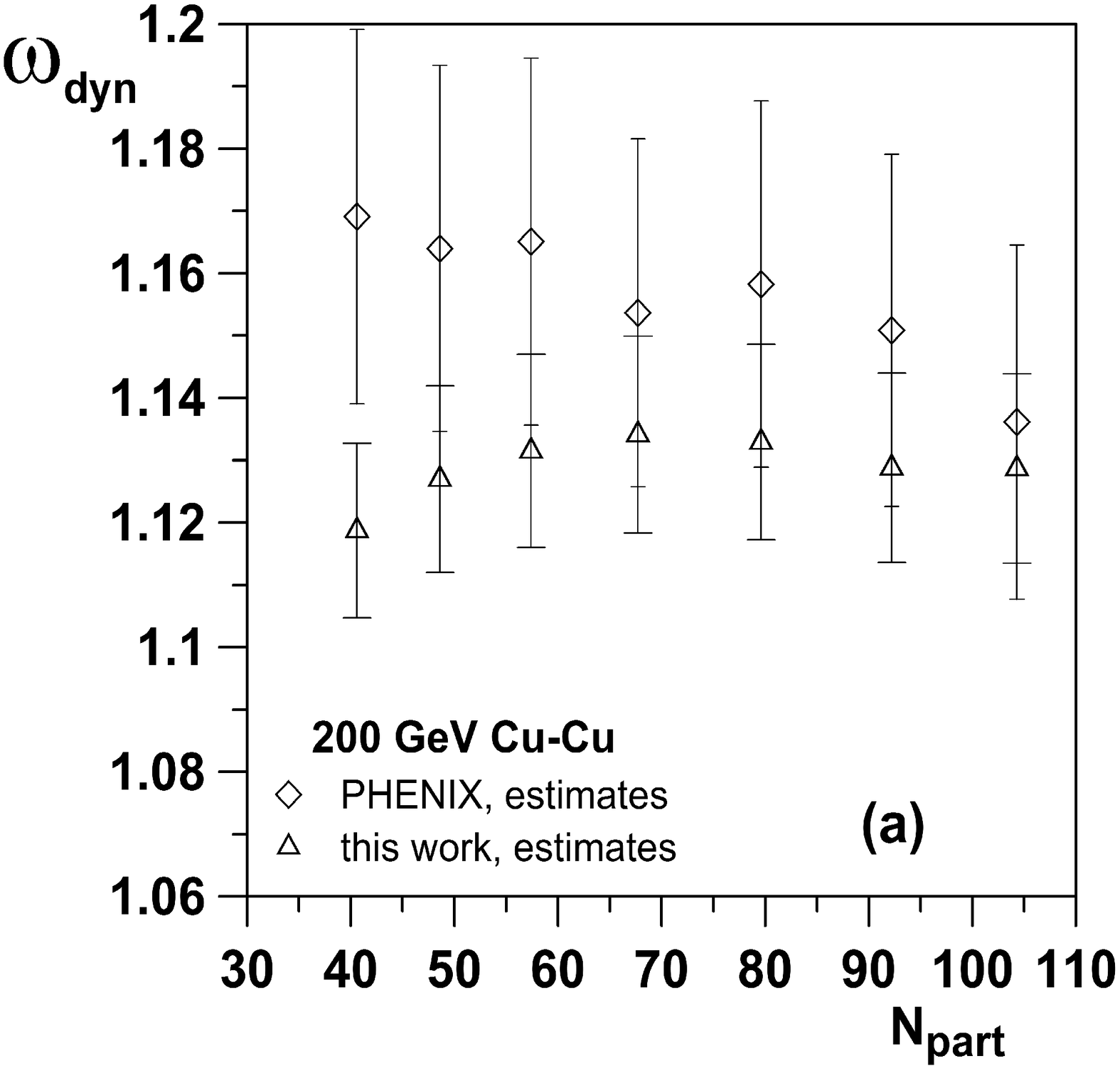}}} & $\;\;\;\;\;\;$
  {\resizebox{!}{8cm}{\includegraphics{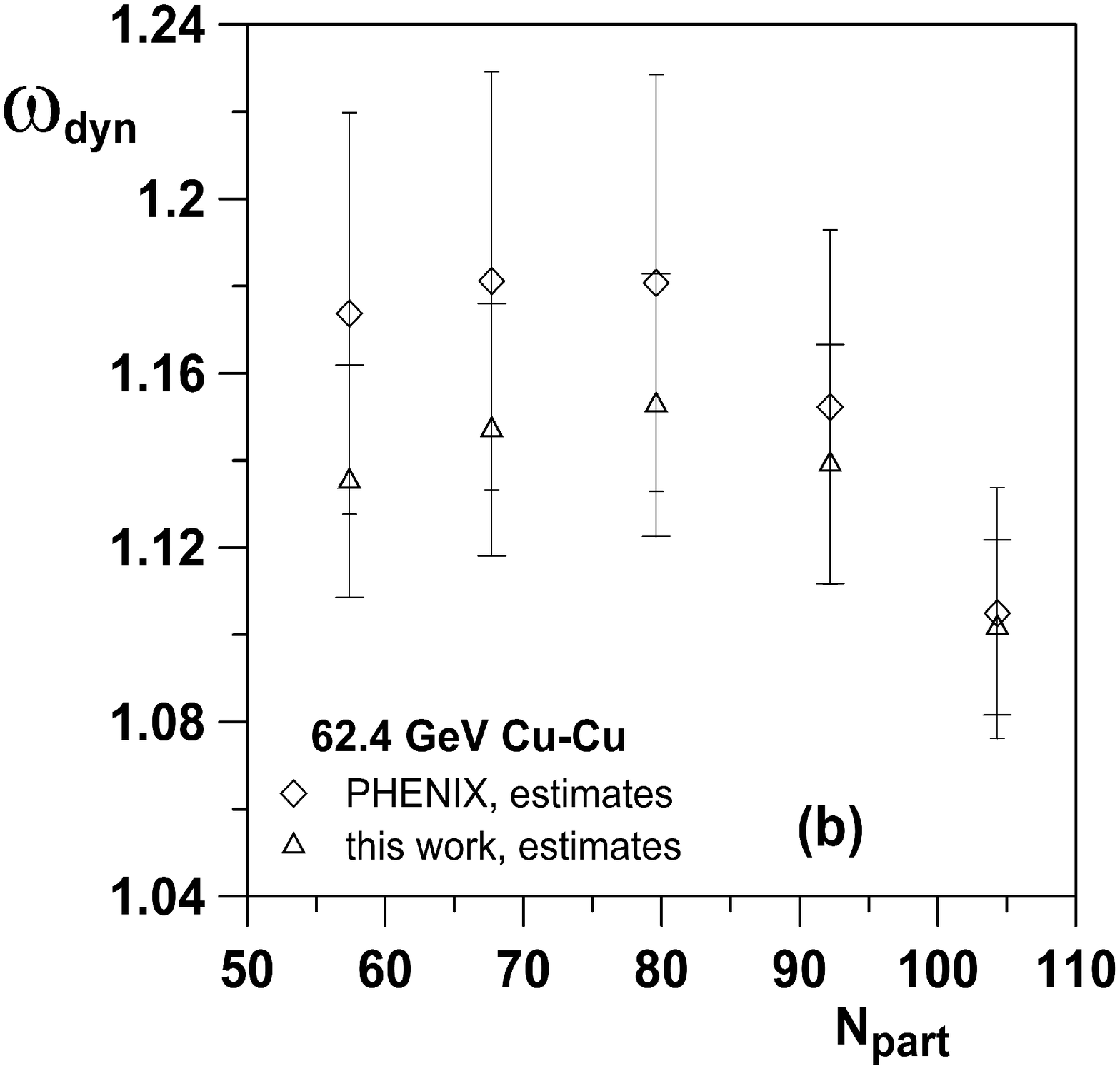}}}
  \end{tabular}
    \caption{ Scaled variance for 200 GeV (a) and 62.4 GeV (b) Cu-Cu collisions. PHENIX estimates are from \protect\cite{Adare:2008ns}. Estimates from this work are for the cases with ranges limited to the bins where $n_{i} > 60$, see Tables~\ref{Table6} and \ref{Table8}.}
  \label{fig8}
 \end{center}
\end{figure}

Another surprising point is the comparison of the values of the PHENIX test statistic $\chi_{PHEN}^{2}$ divided by $n_d$, the ninth column of Tables~\ref{Table1}-\ref{Table8}, with the corresponding values of Ref.~\cite{Adare:2008ns}. For the choice (\textit{ii}), Tables~\ref{Table2}, \ref{Table4}, \ref{Table6} and \ref{Table8}, the $\chi_{PHEN}^2$/$n_d$ values obtained here are lower than corresponding ones in Ref.~\cite{Adare:2008ns}. Values of the parameters $\hat{k}, \hat{\bar{n}}$ are also different from those in Ref.~\cite{Adare:2008ns}, what has resulted in slightly different ($1-3\%$ lower) values of the scaled variance $\omega_{dyn}$, see Figs.~\ref{fig7} and \ref{fig8}. To make the comparison easier also values of $\hat{k}_{dyn}^{-1}$ are presented in the fifth column of Tables~\ref{Table1}-\ref{Table8}. Generally, $\hat{\bar{n}}$ is greater but the difference does not exceed $10\%$ and decreases with the centrality. $\hat{k}_{dyn}^{-1}$ is smaller, especially for case (\textit{ii}) and the difference also decreases with the centrality; from about $20-30\%$ for the least central classes to about $5-10\%$ for the most central ones.


\section{Conclusions}
\label{Conclus}

Results of the likelihood ratio test (likelihood $\chi^2$) suggest that the hypothesis of the NBD of charged-particle multiplicities measured by the PHENIX Collaboration
in Au-Au and Cu-Cu collisions at $\sqrt{s_{NN}} = 62.4$ and 200 GeV
should be rejected for all centrality classes. However, it must be stressed that the maximum likelihood method and the likelihood ratio test do not take actual experimental errors into account. This could be seen as a drawback but, in fact, only the LS test statistic takes actual experimental errors into account. Then the problem with the size of errors might occur when the LS method is used not only to fit parameters of a theoretical model but also to assess how confident the rejection or acceptance of a hypothesis is. This is because too big or too small errors cause the false
inference in this case. But the judgement whether errors are too big already or still adequate is subjective. When errors are large enough it is likely that a false hypothesis would be accepted (this situation is called ''error of the second kind'' in statistics \cite{Cowan:1998ji,James:2006zz,Hoel:1971aa}). Also one can encounter serious difficulties when tries to express somehow the goodness-of-fit when the LS method is applied, as it has been explained in Appendix B.

The goodness-of-fit expressed by the $P$-value is necessary to assess the quality of fit. Here is an example: let $\chi^2/n_d = 1.5$ for a test which is $\chi^2$ distributed. Is this fit good or bad? Well, it depends on $n_d$. But how to find any quantitative measure to decide? This measure is the $P$-value. For $n_d = 10$, $P = 0.13$ so the fit should be accepted at the significance level $0.1 \%$, but for $n_d = 100$, $P = 0.0009$ so the fit should be rejected at the same significance level (Ref.~\cite{Cowan:1998ji}, p.62). But to calculate the $P$-value one has to know the distribution of the test statistic at the parameter estimates. In the general case of the LS test statistic this distribution is unknown, unless very specific assumptions are fulfilled as it has been shown in Appendix B. Certainly, assumptions 1 and 3 are not fulfilled when the NBD hypothesis is tested and systematic errors are added in quadrature to statistical ones. Thus at the beginning of the investigations the situation is the following: the likelihood $\chi^2$ does not take the errors into account, but its distribution is known asymptotically; the LS test statistic takes errors (including systematic ones) into account but its distribution is not known, even asymptotically. In the PHENIX case and with their estimations of systematic errors, these problems have been resolved naturally, i.e. both goals have been achieved - statistical and systematic errors are taken into account and the test statistic distribution is known.

The application of the LS method, in the way as the PHENIX Collaboration did, i.e. with their systematic errors included, has revealed a few very interesting things. First of all it has turned out that the corresponding LS test statistic (the PHENIX test statistic $\chi_{PHEN}^{2}$) equals the Neyman's $\chi^2$ test statistic multiplied by 0.1. This enables to use the well known asymptotic properties of the Neyman's $\chi^2$ to find the asymptotic distribution of the PHENIX test statistic, so the goodness-of-fit can be now calculated because sample sizes are very large here. Additionally, PHENIX test statistic estimators of NBD parameters are Neyman's $\chi^2$ estimators. But likelihood $\chi^2$ and Neyman's $\chi^2$ test statistics are asymptotically equivalent, so for a very large sample their estimators (and estimates) should coincide. Therefore determination of NBD parameters with the use of ML method and then insertion of them into the PHENIX test statistic is reasonable. Note that this way of the determination of NBD parameters has turned out to be much simpler than with the use of the LS method, e.g. the optimal $\bar{n}$ equals $\langle N_{ch} \rangle$ (see Appendix~\ref{Aaa}). And last but not least, because the likelihood $\chi^2$ converges faster to efficiency then the Neyman's $\chi^2$, this method should be preferable when estimation of parameters and errors on estimates are considered (Ref.~\cite{James:2006zz}, p. 193; Ref.~\cite{Kendall:1999bb}, Sec. 18.59).

The correct inference from the results of the PHENIX test statistic $\chi_{PHEN}^{2}$, i.e. the test statistic which in opposite to the likelihood $\chi^2$ takes the systematic errors into account, shows that the hypothesis of the NBD of charged-particle multiplicities measured in Au-Au and Cu-Cu collisions at $\sqrt{s_{NN}} = 62.4$ and 200 GeV should be accepted roughly in one fourth of PHENIX classes of the collision system and centrality. In particular, for the PHENIX collision system Au-Au at $\sqrt{s_{NN}} = 62.4$ GeV as a whole the hypothesis of the NBD could not be rejected, whereas for the Cu-Cu system at the same energy should be rejected. For two other systems (both at $\sqrt{s_{NN}} = 200$ GeV) the hypothesis of the NBD seems to be very unlikely.

\begin{acknowledgments}
The author thanks Jeffery Mitchell for helpful explanations of the PHENIX data.
This work was supported in part by the Polish Ministry of Science and Higher
Education under contract No. N N202 0523 40.
\end{acknowledgments}

\begin{table*}[!]
\caption{\label{Table1} Results of fitting multiplicity
distributions measured by the PHENIX Collaboration in Au-Au
collisions at $\sqrt{s_{NN}} = 200$ GeV,
$f_{geo}= 0.37 \pm 0.027$ \protect\cite{Adare:2008ns}.
Fitting ranges are limited to the bins with $n_{i}
> 5$, where $n_{i}$ is the number of events in the $i$th bin. }
\begin{ruledtabular}
\begin{tabular}{cccccccccc}\hline
 & & & & & & $\chi^2_{\lambda}$/$n_d$ & & &
\\
 Centrality & N & $\hat{k}$ & $\hat{\bar{n}}$ & $1/\hat{k}_{dyn}$ &
 $\omega_{dyn}$ & $\chi^2_{\lambda}$ & P-value & $\chi_{PHEN}^2$/$n_d$ & P-value
\\
  $[\%]$ & & & & & & ($n_d$) & [\%]& $\chi_{PHEN}^2$ & [\%]
\\
\hline
 0-5 & 653145 & 270.0 & 61.85 & 1.37$\;\cdot 10^{-3}$ & 1.08 & 23.73 & 0 & 0.98 & 0
\\
 &  & $\pm 2.5$ & $\pm 0.01$ & $\pm 0.10\cdot\! 10^{-3}$ & $\pm 0.01$ & 1756.0 & & 72.36 &
\\
 &  & & & & & (74) & & &
\\
 & & & & &
\\
 5-10 & 657944 & 163.4 & 53.91 & 2.26$\;\cdot 10^{-3}$ & 1.12 & 9.12 & 0 & 0.69 & 0
\\
 & & $\pm 1.2$ & $\pm 0.01$ & $\pm 0.17\cdot\! 10^{-3}$ & $\pm 0.01$ & 592.7 & & 44.95 &
\\
 & & & & & & (65) & & &
\\
 & & & & &
\\
 10-15 & 658739 & 112.5 & 46.50 & 3.29$\;\cdot 10^{-3}$ & 1.15 & 11.5 & 0 & 0.66 & 0
\\
 & & $\pm 0.7$ & $\pm 0.01$ & $\pm 0.24\cdot\! 10^{-3}$ & $\pm 0.01$ & 795.5 & & 45.43 &
\\
 & & & & & & (69) & & &
\\
 & & & & &
\\
 15-20 & 659607 & 85.1 & 39.72 & 4.35$\;\cdot 10^{-3}$ & 1.17 & 8.9 & 0 & 0.52 & 0
\\
 & & $\pm 0.5$ & $\pm 0.01$ & $\pm 0.32\cdot\! 10^{-3}$ & $\pm 0.01$ & 585.8 & & 34.20 &
\\
 & & & & & & (66) & & &
\\
 & & & & &
\\
 20-25 & 658785 &  67.6 & 33.56 & 5.48$\;\cdot 10^{-3}$ & 1.18 & 13.5 & 0 & 0.46 & 0
\\
 & & $\pm 0.4$ & $\pm 0.01$ & $\pm 0.40\cdot\! 10^{-3}$ & $\pm 0.01$ & 848.8 & & 29.01 &
\\
 & & & & & & (63) & & &
\\
 & & & & &
\\
 25-30 & 659632 & 56.7 & 28.01 & 6.52$\;\cdot 10^{-3}$ & 1.18 & 10.9 & 0 & 0.37 & 0
\\
 & & $\pm 0.3$ & $\pm 0.01$ & $\pm 0.48\cdot\! 10^{-3}$ & $\pm 0.01$ & 640.6 & & 22.10 &
\\
 & & & & & & (59) & & &
\\
 & & & & &
\\
 30-35 & 659303 & 47.4 & 23.02 & 7.81$\;\cdot 10^{-3}$ & 1.18 & 7.9 & 0 & 0.31 & 0
\\
 & & $\pm 0.3$ & $\pm 0.01$ & $\pm 0.57\cdot\! 10^{-3}$ & $\pm 0.01$ & 429.9 & & 16.72 &
\\
 & & & & & & (54) & & &
\\
 & & & & &
\\
 35-40 & 661174 & 40.5 & 18.64 & 9.13$\;\cdot 10^{-3}$ & 1.17 & 8.5 & 0 & 0.37 & 0
\\
 & & $\pm 0.2$ & $\pm 0.01$ & $\pm 0.67\cdot\! 10^{-3}$ & $\pm 0.01$ & 389.7 & & 17.21 &
\\
 & & & & & & (46) & & &
\\
 & & & & &
\\
 40-45 & 661599 & 34.0 & 14.84 & 1.09$\;\cdot 10^{-2}$ & 1.16 & 7.3 & 0 & 0.35 & 0
\\
 & & $\pm 0.2$ & $\pm 0.01$ & $\pm 0.80\cdot\! 10^{-3}$ & $\pm 0.01$ & 301.0 & & 14.34 &
\\
 & & & & & & (41) & & &
\\
 & & & & &
\\
 45-50 & 661765 & 27.3 & 11.57 & 1.35$\;\cdot 10^{-2}$ & 1.16 & 10.5 & 0 & 0.92 & 0
\\
 & & $\pm 0.2$ & $\pm 0.005$ & $\pm 0.99\cdot\! 10^{-3}$ & $\pm 0.01$ & 390.2 & & 34.19 &
\\
 & & & & & & (37) & & &
\\
 & & & & &
\\
 50-55 & 662114 & 21.3 & 8.82 & 1.74$\;\cdot 10^{-2}$ & 1.15 & 38.8 & 0 & 12.06 & 0
\\
 & & $\pm 0.1$ & $\pm 0.004$ & $\pm 0.13\cdot\! 10^{-2}$ & $\pm 0.01$ & 1436.4 & & 446.2 &
\\
 & & & & & & (37) & & &
\\
 & & & & &
\\
\hline
\end{tabular}
\end{ruledtabular}
\end{table*}

\begin{table*}[!]
\caption{\label{Table2} Results of fitting multiplicity
distributions measured by the PHENIX Collaboration in Au-Au
collisions at $\sqrt{s_{NN}} = 200$ GeV,
$f_{geo}= 0.37 \pm 0.027$ \protect\cite{Adare:2008ns}.
Fitting ranges are limited to the bins with $n_{i}
> 60$, where $n_{i}$ is the number of events in the $i$th bin. }
\begin{ruledtabular}
\begin{tabular}{cccccccccc} \hline
 & & & & & &  $\chi^2_{\lambda}$/$n_d$ & & &
\\
 Centrality & N & $\hat{k}$ & $\hat{\bar{n}}$ & $1/\hat{k}_{dyn}$ &
 $\omega_{dyn}$ & $\chi^2_{\lambda}$ & P-value & $\chi_{PHEN}^2$/$n_d$ & P-value
\\
  $[\%]$ & & & & & &  ($n_d$) & [\%] & $\chi_{PHEN}^2$ & [\%]
\\
\hline
 0-5 & 652579 & 289.0 & 61.86 & 1.28$\;\cdot 10^{-3}$ & 1.08 & 20.0 & 0 & 0.57 & 0
\\
 &  & $\pm 2.9$ & $\pm 0.01$ & $\pm 0.94\cdot\! 10^{-4}$ & $\pm 0.01$ & 1160.2 & & 32.86 &
\\
 &  & & & & &  (58) & & &
\\
 & & & & &
\\
 5-10 & 657571 & 168.1 & 53.91 & 2.20$\;\cdot 10^{-3}$ & 1.12 & 20.56 & 0 & 0.61 & 0
\\
 & & $\pm 1.2$ & $\pm 0.01$ & $\pm 0.16\cdot\! 10^{-3}$ & $\pm 0.01$ & 1151.6 & & 34.41 &
\\
 & & & & & &  (56) & & &
\\
 & & & & &
\\
 10-15 & 658258 & 116.4 & 46.50 & 3.18$\;\cdot 10^{-3}$ & 1.15 & 18.4 & 0 & 0.53 & 0
\\
 & & $\pm 0.7$ & $\pm 0.01$ & $\pm 0.23\cdot\! 10^{-3}$ & $\pm 0.01$ & 991.7 & & 28.81 &
\\
 & & & & & &  (54) & & &
\\
 & & & & &
\\
 15-20 & 659302 & 86.9 & 39.72 & 4.26$\;\cdot 10^{-3}$ & 1.17 & 12.6 & 0 & 0.43 & 0
\\
 & & $\pm 0.5$ & $\pm 0.01$ & $\pm 0.31\cdot\! 10^{-3}$ & $\pm 0.01$ & 667.5 & & 22.97 &
\\
 & & & & & &  (53) & & &
\\
 & & & & &
\\
 20-25 & 658461 &  69.1 & 33.56 & 5.36$\;\cdot 10^{-3}$ & 1.18 & 12.3 & 0 & 0.34 & 0
\\
 & & $\pm 0.4$ & $\pm 0.01$ & $\pm 0.39\cdot\! 10^{-3}$ & $\pm 0.01$ & 604.7 & & 16.46 &
\\
 & & & & & &  (49) & & &
\\
 & & & & &
\\
 25-30 & 659337 & 57.9 & 28.0 & 6.39$\;\cdot 10^{-3}$ & 1.18 & 10.4 & 0 & 0.28 & 6.7$\cdot 10^{-8}$
\\
 & & $\pm 0.3$ & $\pm 0.01$ & $\pm 0.47\cdot\! 10^{-3}$ & $\pm 0.01$ & 469.1 & & 12.80 &
\\
 & & & & & &  (45) & & &
\\
 & & & & &
\\
 30-35 & 659021 & 48.3 & 23.02 & 7.66$\;\cdot 10^{-3}$ & 1.18 & 8.6 & 0 & 0.16 & 0.76
\\
 & & $\pm 0.3$ & $\pm 0.01$ & $\pm 0.56\cdot\! 10^{-3}$ & $\pm 0.01$ & 351.02 & & 6.62 &
\\
 & & & & & &  (41) & & &
\\
 & & & & &
\\
 35-40 & 660937 & 41.3 & 18.64 & 8.96$\;\cdot 10^{-3}$ & 1.17 & 7.6 & 0 & 0.19 & 0.12
\\
 & & $\pm 0.2$ & $\pm 0.01$ & $\pm 0.66\cdot\! 10^{-3}$ & $\pm 0.01$ & 280.3 & & 6.85 &
\\
 & & & & & &  (37) & & &
\\
 & & & & &
\\
 40-45 & 661422 & 34.6 & 14.84 & 1.07$\;\cdot 10^{-2}$ & 1.16 & 7.9 & 0 & 0.21 & 0.015
\\
 & & $\pm 0.2$ & $\pm 0.01$ & $\pm 0.78\cdot\! 10^{-3}$ & $\pm 0.01$ & 260.3 & & 7.06 &
\\
 & & & & & &  (33) & & &
\\
 & & & & &
\\
 45-50 & 661577 & 27.9 & 11.56 & 1.33$\;\cdot 10^{-2}$ & 1.15 & 10.0 & 0 & 0.23 & 0.011
\\
 & & $\pm 0.2$ & $\pm 0.005$ & $\pm 0.97\cdot\! 10^{-3}$ & $\pm 0.01$ & 279.9 & & 6.44 &
\\
 & & & & & &  (28) & & &
\\
 & & & & &
\\
 50-55 & 661877 & 21.9 & 8.81 & 1.69$\;\cdot 10^{-2}$ & 1.15 & 40.0 & 0 & 0.30 & 7.8$\cdot 10^{-5}$
\\
 & & $\pm 0.1$ & $\pm 0.004$ & $\pm 0.12\cdot\! 10^{-2}$ & $\pm 0.01$ & 959.2 & & 7.29 &
\\
 & & & & & &  (24) & & &
\\
 & & & & &
\\
\hline
\end{tabular}
\end{ruledtabular}
\end{table*}

\begin{table*}[!]
\caption{\label{Table3} Results of fitting multiplicity
distributions measured by the PHENIX Collaboration in Au-Au
collisions at $\sqrt{s_{NN}} = 62.4$ GeV, $f_{geo}= 0.33 \pm 0.031$
\protect\cite{Adare:2008ns}. Fitting ranges are limited to the bins
with $n_{i} > 5$, where $n_{i}$ is the number of events in the $i$th
bin. }
\begin{ruledtabular}
\begin{tabular}{cccccccccc}\hline
 & & & & & & $\chi^2_{\lambda}$/$n_d$ & & &
\\
 Centrality & N & $\hat{k}$ & $\hat{\bar{n}}$ & $1/\hat{k}_{dyn}$ &
 $\omega_{dyn}$ & $\chi^2_{\lambda}$ & P-value & $\chi_{PHEN}^2$/$n_d$ & P-value
\\
  $[\%]$ & & & & & & ($n_d$) & [\%] & $\chi_{PHEN}^2$ & [\%]
\\
\hline
 0-5 & 607155 & 225.2 & 44.67 & 1.47$\;\cdot 10^{-3}$ & 1.07 & 2.37
 & 1.7$\cdot 10^{-8}$ & 0.18 & 0.015
\\
 &  & $\pm 2.5$ & $\pm 0.01$ & $\pm 0.14\cdot\! 10^{-3}$ & $\pm 0.01$ & 139.6 & & 10.65 &
\\
 &  & & & & & (59) & & &
\\
 & & & & &
\\
 5-10 & 752392 & 142.3 & 37.96 & 2.32$\;\cdot 10^{-3}$ & 1.09 & 2.44
 & 1.9$\cdot 10^{-8}$ & 0.11 & 29.3
\\
 & & $\pm 1.1$ & $\pm 0.01$ & $\pm 0.22\cdot\! 10^{-3}$ & $\pm 0.01$ & 131.9 & & 5.91 &
\\
 & & & & & & (54) & & &
\\
 & & & & &
\\
 10-15 & 752837 & 115.2 & 31.53 & 2.87$\;\cdot 10^{-3}$ & 1.09 & 2.06
 & 1.1$\cdot 10^{-5}$ & 0.13 & 6.0
\\
 & & $\pm 0.9$ & $\pm 0.01$ & $\pm 0.27\cdot\! 10^{-3}$ & $\pm 0.01$ & 107.1 & & 6.88 &
\\
 & & & & & & (52) & & &
\\
 & & & & &
\\
 15-20 & 752553 & 88.0 & 26.07 & 3.75$\;\cdot 10^{-3}$ & 1.10 & 1.86
 & 3.2$\cdot 10^{-4}$ & 0.13 & 9.9
\\
 & & $\pm 0.6$ & $\pm 0.01$ & $\pm 0.35\cdot\! 10^{-3}$ & $\pm 0.01$ & 87.3 & & 5.98 &
\\
 & & & & & & (47) & & &
\\
 & & & & &
\\
 20-25 & 752296 &  68.5 & 21.35 & 4.82$\;\cdot 10^{-3}$ & 1.10 & 2.63
 & 3.1$\cdot 10^{-8}$ & 0.21 & 2.7$\cdot 10^{-3}$
\\
 & & $\pm 0.5$ & $\pm 0.01$ & $\pm 0.45\cdot\! 10^{-3}$ & $\pm 0.01$ & 113.2 & & 9.10 &
\\
 & & & & & & (43) & & &
\\
 & & & & &
\\
 25-30 & 752183 & 53.2 & 17.30 & 6.21$\;\cdot 10^{-3}$ & 1.11 & 2.75
 & 2.7$\cdot 10^{-8}$ & 0.23 & 1.2$\cdot 10^{-3}$
\\
 & & $\pm  0.4$ & $\pm 0.01$ & $\pm 0.59\cdot\! 10^{-3}$ & $\pm 0.01$ & 107.3 & & 8.81 &
\\
 & & & & & & (39) & & &
\\
 & & & & &
\\
 30-35 & 751375 & 40.1 & 13.84 & 8.22$\;\cdot 10^{-3}$ & 1.11 & 2.97
 & 9.6$\cdot 10^{-9}$ & 0.25 & 3.0$\cdot 10^{-4}$
\\
 & & $\pm 0.3$ & $\pm 0.005$ & $\pm 0.77\cdot\! 10^{-3}$ & $\pm 0.01$ & 103.9 & & 8.65 &
\\
 & & & & & & (35) & & &
\\
 & & & & &
\\
 35-40 & 751661 & 31.7 & 10.89 & 1.04$\;\cdot 10^{-2}$ & 1.11 & 6.72 & 0 & 0.16 & 2.7
\\
 & & $\pm 0.2$ & $\pm 0.004$ & $\pm 0.98\cdot\! 10^{-3}$ & $\pm 0.01$ & 194.9 & & 4.54 &
\\
 & & & & & & (29) & & &
\\
 & & & & &
\\
 40-45 & 750884 & 25.1 & 8.42 & 1.31$\;\cdot 10^{-2}$ & 1.11 & 37.5 & 0 & 40.36 & 0
\\
 & & $\pm 0.2$ & $\pm 0.004$ & $\pm 0.12\cdot\! 10^{-2}$ & $\pm 0.01$ & 937.4 & & 1009.1 &
\\
 & & & & & & (25) & & &
\\
 & & & & &
\\
 45-50 & 751421 & 21.8 & 6.41 & 1.51$\;\cdot 10^{-2}$ & 1.10 & 209.0
 & 0 & 285.9 & 0
\\
 & & $\pm 0.2$ & $\pm 0.003$ & $\pm 0.14\cdot\! 10^{-2}$ & $\pm 0.01$ & 4806.8 & & 6576.7 &
\\
 & & & & & & (23) & & &
\\
 & & & & &
\\
\hline
\end{tabular}
\end{ruledtabular}
\end{table*}

\begin{table*}[!]
\caption{\label{Table4} Results of fitting multiplicity
distributions measured by the PHENIX Collaboration in Au-Au
collisions at $\sqrt{s_{NN}} = 62.4$ GeV, $f_{geo}= 0.33 \pm 0.031$
\protect\cite{Adare:2008ns}. Fitting ranges are limited to the bins
with $n_{i} > 40$, where $n_{i}$ is the number of events in the
$i$th bin. }
\begin{ruledtabular}
\begin{tabular}{cccccccccc}\hline
 & & & & & &  $\chi^2_{\lambda}$/$n_d$ & & &
\\
 Centrality & N & $\hat{k}$ & $\hat{\bar{n}}$ & $1/\hat{k}_{dyn}$ &
 $\omega_{dyn}$ & $\chi^2_{\lambda}$ & P-value & $\chi_{PHEN}^2$/$n_d$ & P-value
\\
  $[\%]$ & & & & & &  ($n_d$) & [\%] & $\chi_{PHEN}^2$ & [\%]
\\
\hline
 0-5 & 607075 & 227.9 & 44.67 & 1.45$\;\cdot 10^{-3}$ & 1.06 & 5.55 & 0 & 0.19 & 5.6$\cdot 10^{-3}$
\\
 &  & $\pm 2.5$ & $\pm 0.01$ & $\pm 0.14\cdot\! 10^{-3}$ & $\pm 0.01$ & 294.3 & & 10.2 &
\\
 &  & & & & &  (53) & & &
\\
 & & & & &
\\
 5-10 & 752263 & 143.9 & 37.96 & 2.29$\;\cdot 10^{-3}$ & 1.09 & 7.80 & 0 & 0.12 & 14.4
\\
 & & $\pm 1.1$ & $\pm 0.01$ & $\pm 0.22\cdot\! 10^{-3}$ & $\pm 0.01$ & 382.4 & & 5.95 &
\\
 & & & & & &  (49) & & &
\\
 & & & & &
\\
 10-15 & 752739 & 116.2 & 31.53 & 2.84$\;\cdot 10^{-3}$ & 1.09 & 5.67 & 0 & 0.13 & 7.0
\\
 & & $\pm 0.9$ & $\pm 0.01$ & $\pm 0.27\cdot\! 10^{-3}$ & $\pm 0.01$ & 260.8 & & 6.08 &
\\
 & & & & & &  (46) & & &
\\
 & & & & &
\\
 15-20 & 752492 & 88.5 & 26.07 & 3.73$\;\cdot 10^{-3}$ & 1.10 & 5.97 & 0 & 0.11 & 30.9
\\
 & & $\pm 0.6$ & $\pm 0.01$ & $\pm 0.35\cdot\! 10^{-3}$ & $\pm 0.01$ & 250.9 & & 4.60 &
\\
 & & & & & &  (42) & & &
\\
 & & & & &
\\
 20-25 & 752182 &  69.2 & 21.35 & 4.77$\;\cdot 10^{-3}$ & 1.10 & 10.2 & 0 & 0.22 & 2.4$\cdot 10^{-3}$
\\
 & & $\pm 0.5$ & $\pm 0.01$ & $\pm 0.45\cdot\! 10^{-3}$ & $\pm 0.01$ & 377.2 & & 8.27 &
\\
 & & & & & &  (37) & & &
\\
 & & & & &
\\
 25-30 & 752095 & 53.6 & 17.30 & 6.16$\;\cdot 10^{-3}$ & 1.11 & 8.2 & 0 & 0.23 & 1.8$\cdot 10^{-3}$
\\
 & & $\pm  0.4$ & $\pm 0.01$ & $\pm 0.58\cdot\! 10^{-3}$ & $\pm 0.01$ & 279.2 & & 7.92 &
\\
 & & & & & &  (34) & & &
\\
 & & & & &
\\
 30-35 & 751324 & 40.3 & 13.84 & 8.19$\;\cdot 10^{-3}$ & 1.11 & 7.40 & 0 & 0.26 & 4.3$\cdot 10^{-4}$
\\
 & & $\pm 0.3$ & $\pm 0.005$ & $\pm 0.77\cdot\! 10^{-3}$ & $\pm 0.01$ & 229.3 & & 7.92 &
\\
 & & & & & &  (31) & & &
\\
 & & & & &
\\
 35-40 & 751639 & 31.8 & 10.89 & 1.04$\;\cdot 10^{-2}$ & 1.11 & 9.43 & 0 & 0.15 & 3.5
\\
 & & $\pm 0.2$ & $\pm 0.004$ & $\pm 0.98\cdot\! 10^{-3}$ & $\pm 0.01$ & 254.7 & & 4.17 &
\\
 & & & & & &  (27) & & &
\\
 & & & & &
\\
 40-45 & 750852 & 25.2 & 8.42 & 1.31$\;\cdot 10^{-2}$ & 1.11 & 50.7 & 0 & 0.22 & 0.062
\\
 & & $\pm 0.2$ & $\pm 0.004$ & $\pm 0.12\cdot\! 10^{-2}$ & $\pm 0.01$ & 1166.3 & & 5.13 &
\\
 & & & & & &  (23) & & &
\\
 & & & & &
\\
 45-50 & 751348 & 22.0 & 6.41 & 1.50$\;\cdot 10^{-2}$ & 1.10 & 259.8
 & 0 & 343.1 & 0
\\
 & & $\pm 0.2$ & $\pm 0.003$ & $\pm 0.14\cdot\! 10^{-2}$ & $\pm 0.01$ & 4936.4 & & 6519.1 &
\\
 & & & & & &  (19) & & &
\\
 & & & & &
\\
\hline
\end{tabular}
\end{ruledtabular}
\end{table*}

\begin{table*}[!]
\caption{\label{Table5} Results of fitting multiplicity
distributions measured by the PHENIX Collaboration in Cu-Cu
collisions at $\sqrt{s_{NN}} = 200$ GeV,
$f_{geo}= 0.40 \pm 0.047$ \protect\cite{Adare:2008ns}.
Fitting ranges are limited to the bins with $n_{i}
> 5$, where $n_{i}$ is the number of events in the $i$th bin. }
\begin{ruledtabular}
\begin{tabular}{cccccccccc}\hline
 & & & & & &  $\chi^2_{\lambda}$/$n_d$ & & &
\\
 Centrality & N & $\hat{k}$ & $\hat{\bar{n}}$ & $1/\hat{k}_{dyn}$ &
 $\omega_{dyn}$ & $\chi^2_{\lambda}$ & P-value & $\chi_{PHEN}^2$/$n_d$ & P-value
\\
  $[\%]$ & & & & & &  ($n_d$) & [\%] & $\chi_{PHEN}^2$ & [\%]
\\
\hline
 0-5 & 368510 & 59.6 & 19.80 & 6.72$\;\cdot 10^{-3}$ & 1.13 & 94.8 & 0 & 2.1 & 0
\\
 &  & $\pm 0.6$ & $\pm 0.01$ & $\pm 0.79\cdot\! 10^{-3}$ & $\pm 0.02$ & 3887.0 & & 87.1 &
\\
 &  & & & & &  (41) & & &
\\
 & & & & &
\\
 5-10 & 369206 & 49.6 & 16.74 & 8.06$\;\cdot 10^{-3}$ & 1.13 & 16.5 & 0 & 0.66 & 0
\\
 & & $\pm 0.5$ & $\pm 0.01$ & $\pm 0.95\cdot\! 10^{-3}$ & $\pm 0.02$ & 628.5 & & 25.3 &
\\
 & & & & & &  (38) & & &
\\
 & & & & &
\\
 10-15 & 369945 & 41.5 & 14.05 & 9.64$\;\cdot 10^{-3}$ & 1.14 & 6.8 & 0 & 0.38 & 0
\\
 & & $\pm 0.4$ & $\pm 0.01$ & $\pm 0.11\cdot\! 10^{-2}$ & $\pm 0.02$ & 225.5 & & 12.6 &
\\
 & & & & & &  (33) & & &
\\
 & & & & &
\\
 15-20 & 370066 & 34.5 & 11.78 & 1.16$\;\cdot 10^{-2}$ & 1.14 & 3.0
 & 5.8$\cdot 10^{-8}$ & 0.24 & 1.5$\cdot 10^{-3}$
\\
 & & $\pm 0.3$ & $\pm 0.01$ & $\pm 0.14\cdot\! 10^{-2}$ & $\pm 0.02$ & 92.0 & & 7.53 &
\\
 & & & & & &  (31) & & &
\\
 & & & & &
\\
 20-25 & 371877 &  29.2 & 9.81 & 1.37$\;\cdot 10^{-2}$ & 1.13 & 6.6 & 0 & 3.4 & 0
\\
 & & $\pm 0.3$ & $\pm 0.01$ & $\pm 0.16\cdot\! 10^{-2}$ & $\pm 0.02$ & 186.0 & & 93.9 &
\\
 & & & & & &  (28) & & &
\\
 & & & & &
\\
 25-30 & 368876 & 24.9 & 8.14 & 1.60$\;\cdot 10^{-2}$ & 1.13 & 19.3 & 0 & 11.5 & 0
\\
 & & $\pm 0.2$ & $\pm 0.01$ & $\pm 0.19\cdot\! 10^{-2}$ & $\pm 0.02$ & 502.4 & & 298.9 &
\\
 & & & & & &  (26) & & &
\\
 & & & & &
\\
 30-35 & 368072 & 21.9 & 6.72 & 1.83$\;\cdot 10^{-2}$ & 1.12 & 65.6 & 0 & 42.3 & 0
\\
 & & $\pm 0.2$ & $\pm 0.005$ & $\pm 0.22\cdot\! 10^{-2}$ & $\pm 0.01$ & 1704.8 & & 1098.5 &
\\
 & & & & & &  (26) & & &
\\
 & & & & &
\\
\hline
\end{tabular}
\end{ruledtabular}
\end{table*}

\begin{table*}[!]
\caption{\label{Table6} Results of fitting multiplicity
distributions measured by the PHENIX Collaboration in Cu-Cu
collisions at $\sqrt{s_{NN}} = 200$ GeV,
$f_{geo}= 0.40 \pm 0.047$ \protect\cite{Adare:2008ns}.
Fitting ranges are limited to the bins with $n_{i}
> 80$, where $n_{i}$ is the number of events in the $i$th bin. }
\begin{ruledtabular}
\begin{tabular}{cccccccccc}\hline
 & & & & & &  $\chi^2_{\lambda}$/$n_d$ & & &
\\
 Centrality & N & $\hat{k}$ & $\hat{\bar{n}}$ & $1/\hat{k}_{dyn}$ &
 $\omega_{dyn}$ & $\chi^2_{\lambda}$ & P-value &
 $\chi_{PHEN}^2$/$n_d$ & P-value
\\
  $[\%]$ & & & & & &  ($n_d$) & [\%] & $\chi_{PHEN}^2$ & [\%]
\\
\hline
 0-5 & 368271 & 61.5 & 19.79 & 6.50$\;\cdot 10^{-3}$ & 1.13 & 122.2 & 0 & 2.3 & 0
\\
 &  & $\pm 0.6$ & $\pm 0.01$ & $\pm 0.77\cdot\! 10^{-3}$ & $\pm 0.02$ & 4398.3 & & 82.7 &
\\
 &  & & & & &  (36) & & &
\\
 & & & & &
\\
 5-10 & 368869 & 52.0 & 16.74 & 7.69$\;\cdot 10^{-3}$ & 1.13 & 20.5 & 0 & 0.39 & 0
\\
 & & $\pm 0.5$ & $\pm 0.01$ & $\pm 0.91\cdot\! 10^{-3}$ & $\pm 0.02$ & 613.9 & & 11.7 &
\\
 & & & & & &  (30) & & &
\\
 & & & & &
\\
 10-15 & 369825 & 42.3 & 14.05 & 9.46$\;\cdot 10^{-3}$ & 1.13 & 16.2 & 0 & 0.43 & 0
\\
 & & $\pm 0.4$ & $\pm 0.01$ & $\pm 0.11\cdot\! 10^{-2}$ & $\pm 0.02$ & 470.9 & & 12.6 &
\\
 & & & & & &  (29) & & &
\\
 & & & & &
\\
 15-20 & 369964 & 35.1 & 11.77 & 1.14$\;\cdot 10^{-2}$ & 1.13 & 11.4 & 0 & 0.24 & 5.4$\cdot 10^{-3}$
\\
 & & $\pm 0.3$ & $\pm 0.01$ & $\pm 0.13\cdot\! 10^{-2}$ & $\pm 0.02$ & 296.8 & & 6.36 &
\\
 & & & & & &  (26) & & &
\\
 & & & & &
\\
 20-25 & 371752 &  29.8 & 9.80 & 1.34$\;\cdot 10^{-2}$ & 1.13 & 16.1 & 0 & 0.20 & 0.38
\\
 & & $\pm 0.3$ & $\pm 0.01$ & $\pm 0.16\cdot\! 10^{-2}$ & $\pm 0.02$ & 370.4 & & 4.51 &
\\
 & & & & & &  (23) & & &
\\
 & & & & &
\\
 25-30 & 368708 & 25.6 & 8.14 & 1.56$\;\cdot 10^{-2}$ & 1.13 & 42.7 & 0 & 0.21 & 0.23
\\
 & & $\pm 0.3$ & $\pm 0.01$ & $\pm 0.18\cdot\! 10^{-2}$ & $\pm 0.01$ & 853.2 & & 4.27 &
\\
 & & & & & &  (20) & & &
\\
 & & & & &
\\
 30-35 & 367869 & 22.6 & 6.72 & 1.77$\;\cdot 10^{-2}$ & 1.12 & 126.4 & 0 & 0.62 & 0
\\
 & & $\pm 0.2$ & $\pm 0.005$ & $\pm 0.21\cdot\! 10^{-2}$ & $\pm 0.01$ & 2274.4 & & 11.1 &
\\
 & & & & & &  (18) & & &
\\
 & & & & &
\\
\hline
\end{tabular}
\end{ruledtabular}
\end{table*}

\begin{table*}[!]
\caption{\label{Table7} Results of fitting multiplicity
distributions measured by the PHENIX Collaboration in Cu-Cu
collisions at $\sqrt{s_{NN}} = 62.4$ GeV, $f_{geo}= 0.32 \pm 0.063$
\protect\cite{Adare:2008ns}. Fitting ranges are limited to the bins
with $n_{i} > 5$, where $n_{i}$ is the number of events in the $i$th
bin. }
\begin{ruledtabular}
\begin{tabular}{cccccccccc}\hline
 & & & & & &  $\chi^2_{\lambda}$/$n_d$ & & &
\\
 Centrality & N & $\hat{k}$ & $\hat{\bar{n}}$ & $1/\hat{k}_{dyn}$ &
 $\omega_{dyn}$ & $\chi^2_{\lambda}$ & P-value &
 $\chi_{PHEN}^2$/$n_d$ & P-value
\\
  $[\%]$ & & & & & &  ($n_d$) & [\%] & $\chi_{PHEN}^2$ & [\%]
\\
\hline
 0-5 & 298182 & 41.6 & 13.35 & 7.69$\;\cdot 10^{-3}$ & 1.10 & 9.3 & 0 & 0.65 & 0
\\
 &  & $\pm 0.4$ & $\pm 0.01$ & $\pm 0.15\cdot\! 10^{-2}$ & $\pm 0.02$ & 279.9 & & 19.4 &
\\
 &  & & & & &  (30) & & &
\\
 & & & & &
\\
 5-10 & 307150 & 26.5 & 11.67 & 1.21$\;\cdot 10^{-2}$ & 1.14 & 9.7 & 0 & 0.78 & 0
\\
 & & $\pm 0.2$ & $\pm 0.01$ & $\pm 0.24\cdot\! 10^{-2}$ & $\pm 0.03$ & 290.7 & & 23.3 &
\\
 & & & & & &  (30) & & &
\\
 & & & & &
\\
 10-15 & 309874 & 20.5 & 9.90 & 1.56$\;\cdot 10^{-2}$ & 1.15 & 9.3 & 0 & 4.4 & 0
\\
 & & $\pm 0.2$ & $\pm 0.01$ & $\pm 0.31\cdot\! 10^{-2}$ & $\pm 0.03$ & 261.1 & & 122.5 &
\\
 & & & & & &  (28) & & &
\\
 & & & & &
\\
 15-20 & 312530 & 17.8 & 8.27 & 1.80$\;\cdot 10^{-2}$ & 1.15 & 26.0 & 0 & 31.6 & 0
\\
 & & $\pm 0.1$ & $\pm 0.01$ & $\pm 0.36\cdot\! 10^{-2}$ & $\pm 0.03$ & 677.1 & & 821.7 &
\\
 & & & & & &  (26) & & &
\\
 & & & & &
\\
 20-25 & 312884 &  16.0 & 6.89 & 1.99$\;\cdot 10^{-2}$ & 1.14 & 75.8 & 0 & 80.9 & 0
\\
 & & $\pm 0.1$ & $\pm 0.01$ & $\pm 0.39\cdot\! 10^{-2}$ & $\pm 0.03$ & 1744.0 & & 1861.4 &
\\
 & & & & & &  (23) & & &
\\
 & & & & &
\\
\hline
\end{tabular}
\end{ruledtabular}
\end{table*}

\begin{table*}[!]
\caption{\label{Table8} Results of fitting multiplicity
distributions measured by the PHENIX Collaboration in Cu-Cu
collisions at $\sqrt{s_{NN}} = 62.4$ GeV, $f_{geo}= 0.32 \pm 0.063$
\protect\cite{Adare:2008ns}. Fitting ranges are limited to the bins
with $n_{i} > 60$, where $n_{i}$ is the number of events in the
$i$th bin. }
\begin{ruledtabular}
\begin{tabular}{cccccccccc}\hline
 & & & & & &  $\chi^2_{\lambda}$/$n_d$ & & &
\\
 Centrality & N & $\hat{k}$ & $\hat{\bar{n}}$ & $1/\hat{k}_{dyn}$ &
 $\omega_{dyn}$ & $\chi^2_{\lambda}$ & P-value &
 $\chi_{PHEN}^2$/$n_d$ & P-value
\\
  $[\%]$ & & & & & &  ($n_d$) & [\%] & $\chi_{PHEN}^2$ & [\%]
\\
\hline
 0-5 & 298131 & 42.0 & 13.35 & 7.62$\;\cdot 10^{-3}$ & 1.10 & 14.7 & 0 & 0.67 & 0
\\
 &  & $\pm 0.5$ & $\pm 0.01$ & $\pm 0.15\cdot\! 10^{-2}$ & $\pm 0.02$ & 411.9 & & 18.9 &
\\
 &  & & & & &  (28) & & &
\\
 & & & & &
\\
 5-10 & 307061 & 26.8 & 11.66 & 1.19$\;\cdot 10^{-2}$ & 1.14 & 19.7 & 0 & 0.86 & 0
\\
 & & $\pm 0.2$ & $\pm 0.01$ & $\pm 0.24\cdot\! 10^{-2}$ & $\pm 0.03$ & 512.5 & & 22.5 &
\\
 & & & & & &  (26) & & &
\\
 & & & & &
\\
 10-15 & 309798 & 20.7 & 9.90 & 1.54$\;\cdot 10^{-2}$ & 1.15 & 19.4 & 0 & 0.38 & 1.1$\cdot 10^{-7}$
\\
 & & $\pm 0.2$ & $\pm 0.01$ & $\pm 0.30\cdot\! 10^{-2}$ & $\pm 0.03$ & 465.5 & & 9.08 &
\\
 & & & & & &  (24) & & &
\\
 & & & & &
\\
 15-20 & 312434 &  18.0 & 8.27 & 1.78$\;\cdot 10^{-2}$ & 1.15 & 46.5 & 0 & 0.40 & 1.9$\cdot 10^{-7}$
\\
 & & $\pm 0.1$ & $\pm 0.01$ & $\pm 0.35\cdot\! 10^{-2}$ & $\pm 0.03$ & 976.4 & & 8.37 &
\\
 & & & & & &  (21) & & &
\\
 & & & & &
\\
 20-25 & 312758 &  16.3 & 6.89 & 1.96$\;\cdot 10^{-2}$ & 1.14 & 118.1 & 0 & 0.63 & 0
\\
 & & $\pm 0.1$ & $\pm 0.01$ & $\pm 0.39\cdot\! 10^{-2}$ & $\pm 0.03$ & 2243.4 & & 12.05 &
\\
 & & & & & &  (19) & & &
\\
 & & & & &
\\
\hline
\end{tabular}
\end{ruledtabular}
\end{table*}

\appendix

\section{}
\label{Aaa}

Dropping terms not depending on the parameters in Eq.~(\ref{Logratfreq}),
one obtains the following form for the log-likelihood function under
consideration:

\begin{equation}
\ln L(\textbf{Y} \mid p, k) = N \sum_{i=1}^{m}\;P_i^{ex}\; \ln
P(Y_i; p, k) \;. \label{LogLikfubi}
\end{equation}
Since the logarithm of the NBD is given by

\begin{eqnarray}
&&\ln P(n; p, k) \cr \cr & &= \sum_{j=1}^{n} \ln{(k+j-1)} + n
\ln{(1-p)} + k \ln{p} - \ln{(n!)}\;, \cr & & \label{LogNBD}
\end{eqnarray}
the necessary conditions for the existence of the maximum have the following form:

\begin{eqnarray}
&&\frac{\partial}{\partial p}\ln L(\textbf{Y} \mid p, k) \cr \cr & &
= N \sum_{i=1}^{m}\;P_i^{ex}\;\bigg [ -Y_i \frac{1}{1-p} +
\frac{k}{p} \bigg ] \cr \cr & & = N \bigg [ -\frac{1}{1-p}
\sum_{i=1}^{m}\;P_i^{ex} Y_i  + \frac{k}{p} \sum_{i=1}^{m}\;P_i^{ex}
\bigg ] \cr \cr & & = N \bigg [ -\frac{1}{1-p} \langle N_{ch}
\rangle + \frac{k}{p} \bigg ] = 0 \;, \label{DlogLikeqp}
\end{eqnarray}
\begin{eqnarray}
&&\frac{\partial}{\partial k}\ln L(\textbf{Y} \mid p, k) \cr \cr & &
= N \sum_{i=1}^{m}\;P_i^{ex}\;\bigg [ \sum_{j=1}^{Y_i}\;
\frac{1}{k+j-1} + \ln{p} \bigg ] \cr \cr & & = N \bigg [
\sum_{i=1}^{m}\;P_i^{ex}\; \sum_{j=1}^{Y_i}\; \frac{1}{k+j-1} +
\ln{p} \bigg ] = 0 \;, \label{DlogLikeqk}
\end{eqnarray}
where the sum over $j$ is 0 if $Y_i = 0$.

From Eqs.~(\ref{DlogLikeqp}) and (\ref{Parametpk}) one can obtain:

\begin{equation}
\langle N_{ch} \rangle = \frac{k(1-p)}{p} = \bar{n}\;.
\label{Peqalaver}
\end{equation}
Expressing $p$ as a function of $k$ and $\langle N_{ch} \rangle$

\begin{equation}
\frac{1}{p} = \frac{\langle N_{ch} \rangle}{k} + 1 \;,
\label{Oneoverp}
\end{equation}
and substituting it to Eq.~(\ref{DlogLikeqk}) the equation which
determines $\hat{k}$ is obtained:

\begin{eqnarray}
&&\frac{\partial}{\partial k}\ln L(\textbf{Y} \mid p, k) \cr \cr & &
= N \bigg [ \sum_{i=1}^{m}\;P_i^{ex}\; \sum_{j=1}^{Y_i}\;
\frac{1}{k+j-1} - \ln{\bigg (1 + \frac{\langle N_{ch} \rangle}{k}
\bigg )} \bigg ] = 0 \;. \cr & &  \label{DlogLikfk}
\end{eqnarray}
The above equation can be solved numerically. Having obtained
$\hat{k}$ and substituting it into Eq.~(\ref{Oneoverp}) $\hat{p}$ is
derived.

\section{Statistical inference in a capsule}
\label{Capsule}

Let $\{Y_1,Y_2,...,Y_N \}$ be a set of repeated observations of a random variable $Y$ or a set of a single observation of $N$-dimensional random variable $\vec{Y} = (Y_1,Y_2,...,Y_N)$ (this appendix is a brief summary based on Refs.~\cite{Cowan:1998ji,James:2006zz}). The null hypothesis, $H_0$, specifies a p.d.f. of $Y$ or a joint p.d.f. of $\vec{Y}$. The test statistic $t$ is a function of the observations (a function of $N$ random variables equivalently): $t = t(Y_1,Y_2,...,Y_N)$. For simplicity let us assume that $t$ is a scalar function. Let $g(t\mid H_0)$ be a given p.d.f. for the statistic $t$ if $H_0$ is true. The qualitative assessment about the compatibility of $H_0$ with the data is expressed as a decision to accept or reject the null hypothesis. This is done by choosing a value $t_{cut}$ , called the cut or decision boundary. Then, for given observations $\{Y_1,Y_2,...,Y_N \}$ $t_{O} = t(Y_1,Y_2,...,Y_N)$ and if $t_{O} > t_{cut}$ , the hypothesis is rejected; if $t_{O} \leq t_{cut}$ , $H_0$ is accepted. Usually  $t_{cut}$ is chosen in such a way that one obtains the assumed probability $\alpha$ to reject $H_0$ if $H_0$ is true - this is called the significance level:

\begin{equation}
\alpha = \int_{t_{cut}}^{\infty} g(t\mid H_0)dt \;.
\label{Signiflev}
\end{equation}

Now, let $\vec{Y}$ be an $N$-dimensional Gaussian random variable with known covariance matrix $V$ but not known expectation values. $\vec{Y}$ is related to another variable $\vec{X}$ in such a way that there is a true value function ($\equiv$ a hypothesis) $\Lambda = \Lambda(X;\vec{\theta})$, which depends on unknown parameters $\vec{\theta} = (\theta_1,...,\theta_m)$ and expectation value of $Y_i$, $E[Y_i] = \Lambda(X_i;\vec{\theta})$. Then one defines the least-squares (LS) statistic as

\begin{equation}
\chi_{LS}^{2}(\vec{Y};\vec{\theta}) = \sum_{i,j=1}^{N} (Y_i-\Lambda(X_i;\vec{\theta}
)) [V^{-1}]_{ij} (Y_j - \Lambda(X_j;\vec{\theta})) \;. \label{ChiLSdef}
\end{equation}
Instead, if one has $N$ independent Gaussian random variables with different unknown means but known variances $\sigma_i^2$ and the true value function $\Lambda = \Lambda(X;\vec{\theta})$, then the LS statistic, Eq.~(\ref{ChiLSdef}), becomes

\begin{equation}
\chi_{LS}^{2}(\vec{Y};\vec{\theta}) = \sum_{i=1}^{N} \frac{(Y_i-\Lambda(X_i;\vec{\theta}))^2}{\sigma_i^2}  \;. \label{ChiLSdef2}
\end{equation}
Let $\vec{Y}$ be a single measurement of the $N$-dimensional random variable (or a set of independent measurements of $N$ random variables) at points $X_1,\;X_2,\;...,\;X_N$. Having replaced the variables by their measured values in Eq.~(\ref{ChiLSdef}) (or Eq.~(\ref{ChiLSdef2})) one converts the LS statistic $\chi_{LS}^{2}(\vec{Y};\vec{\theta})$ into the function of $\vec{\theta}$ only.
The next step is to minimize this function with respect to $\vec{\theta}$. Values of parameters at the minimum are called the LS estimators, $(\hat{\theta}_1,...,\hat{\theta}_m)$. When one has replaced parameters $\vec{\theta}$ (treated as free until now) by their estimators in Eq.~(\ref{ChiLSdef}) (or Eq.~(\ref{ChiLSdef2})), then a test statistic $t_{\chi^{2}} = t_{\chi^{2}} (Y_1,Y_2,...,Y_N) \equiv \chi_{LS,min}^{2}(\vec{Y}) = \chi_{LS}^{2}(\vec{Y}; \hat{\theta}_1,...,\hat{\theta}_m)$ is obtained. What is the decision boundary $t_{\chi^{2},cut}$ for this test statistic? The choice of the proper $t_{\chi^{2},cut}$ is the consequence of the following theorem (see Ref.~\cite{Cowan:1998ji}, pp. 95-96, 104; Ref.~\cite{Frodesen:1979fy}, $\S$10.4.3).
\newline If
\begin{enumerate}
 \item $(Y_1,Y_2,...,Y_N)$ is an $N$-dimensional Gaussian random variable with known covariance matrix $V$ or $(Y_1,Y_2,...,Y_N)$ are independent Gaussian random variables with known variances $\sigma_i^2$;
 \item variables $(X_1,X_2,...,X_N)$ are measured with infinite precision, i.e. without any errors;

 \item the hypothesis $\Lambda(X;\theta_1,...,\theta_m)$ is linear in the parameters $\theta_i$; and

 \item the hypothesis is correct,
\end{enumerate}
then the test statistic $\chi_{LS,min}^{2}$ is distributed according to a $\chi^{2}$ distribution with $n_d = N-m$ degrees of freedom.
\newline If the hypothesis $\Lambda(X;\theta_1,...,\theta_m)$ is nonlinear in the parameters, the exact distribution of $\chi_{LS,min}^{2}$ is not known. However, asymptotically (when $N \longrightarrow \infty$) the distribution of $\chi_{LS,min}^{2}$ approaches a $\chi^{2}$ distribution as well (Ref.~\cite{Frodesen:1979fy}, p. 287; Ref.~\cite{Roe:1992zz}, p. 147). Thus when assumptions 1, 2 and 4 at least are fulfilled and the sample size is large one can consider $\chi_{LS,min}^{2}$ test statistic as $\chi^{2}$ distributed. The expectation value of a random variable $Z$ distributed according to the $\chi^{2}$ distribution with $n_d$ degrees of freedom is $E[Z] = n_d$ and the variance $V[Z] = 2n_d$. As a result 'one expects in a "reasonable" experiment to obtain $\chi_{LS,min}^{2} \approx n_d$' (Ref.~\cite{Beringer:1900zz}, p. 15). Therefore for the test statistic $t_{\chi^{2}} = \chi_{LS,min}^{2}$ the decision boundary $t_{\chi^{2},cut} = E[\chi_{LS,min}^{2}] = n_d$ is chosen. Usually the so-called 'reduced $\chi^{2}$' is reported, which equals $\chi_{LS,min}^{2}/n_d$. So for $\chi_{LS,min}^{2}/n_d$ the decision boundary is just one. It must be stressed here that this choice is the consequence of the fact that the $\chi_{LS,min}^{2}$ test statistic is $\chi^{2}$ distributed. If the distribution of $\chi_{LS,min}^{2}$ is not known at all (e.g. one of the assumptions 1, 2 or 4 is not fulfilled or the sample size is small), this choice is arbitrary - based on common believe rather than on any justification.

The comparison of the actually obtained value of the test statistic $t_{O} = t(Y_1,Y_2,...,Y_N)$ with the decision boundary $t_{cut}$ gives only qualitative information about validity of the hypothesis $H_0$. If one wants to express quantitatively how the null hypothesis agrees with the data a test of goodness-of-fit is necessary \cite{Cowan:1998ji,James:2006zz}. The value of this test shows the level of the compatibility of the observed data with the predictions of $H_0$. This value is given by the probability $P$, under assumption that $H_0$ is true and the experiment would be repeated many times under the same circumstances, of obtaining results as compatible or less with $H_0$ than the result just observed. This probability is called the $P$-value of the test and can be expressed as (Ref.~\cite{James:2006zz}, p. 300)

\begin{equation}
P = \int_{\vec{Y}:t \geq t_{O}} f(\vec{Y} \mid H_0) \;,
\label{Pvaluegen}
\end{equation}
%
where $f(\vec{Y} \mid H_0)$ is the p.d.f. of the $N$-dimensional random variable $\vec{Y}$ under the null hypothesis $H_0$. In general the above integral could be very difficult to calculate unless the p.d.f. $g(t\mid H_0)$ of the test statistic $t$ is known somehow, then one obtains (Ref.~\cite{Beringer:1900zz}, p. 13):

\begin{equation}
P = \int_{t_{O}}^{\infty} g(t\mid H_0)dt \;.
\label{Pvalueg}
\end{equation}
%
Note that this is not the same as Eq.~(\ref{Signiflev}) because that expression is the equation for $t_{cut}$ given the significance level $\alpha$ and should be solved before the measurement, whereas Eq.~(\ref{Pvalueg}) is calculated after the measurement and reflects the obtained (dis)agreement of the observation with the hypothesis $H_0$. The criterion for the rejection or acceptance of $H_0$ can be now formulated with the use of $P$ and $\alpha$ instead of $t_{O}$ and $t_{cut}$: if $P \leq \alpha$ then the hypothesis should be rejected, otherwise should be accepted.

However, the most interesting class of test statistics is such that their distributions are known independently of $H_0$. The most important class consists of so-called '$\chi^{2}$ statistics', i.e. test statistics which are distributed (at least asymptotically) in the $\chi^{2}$ distribution \cite{James:2006zz,Baker:1983tu}. Note that $\chi_{LS}^{2}$ defined earlier, when the assumptions of the theorem are fulfilled, belongs to this class. The likelihood $\chi^2$, Eq.~(\ref{PoissonChi}), the Pearson's $\chi^2$ and the Neyman's $\chi^2$ mentioned in Sec.~\ref{Finl} do as well. Then $P$-value is given by

\begin{equation}
P = \int_{t_{O}}^{\infty}\;f(z;n_d) dz \;, \label{Pvaluechi2}
\end{equation}
where $f(z;n_d)$ is the $\chi^{2}$ p.d.f. and $n_d$ the number of
degrees of freedom.

\section{Wilks's theorem}
\label{Wilks}

Let $X$ be a random variable with p.d.f $f(X,\theta)$, which depends
on parameters $\theta = \{ \theta_1,\; \theta_2,...,\theta_d \} \in
\Theta$, where a parameter space $\Theta$ is an open set in
$\textrm{R}^d$. For the set of $N$ independent observations of $X$,
$\textbf{X} = \{ X_1,\;X_2,...,X_N \}$, one can defined the
likelihood function

\begin{equation}
L(\textbf{X} \mid \theta) = \prod_{j=1}^{N}\; f(X_j; \theta) \;.
\label{LikelfunB}
\end{equation}
%
Now consider $H_0$, a $k$-dimensional subset of $\Theta$,  $k < d$.
Then the maximum likelihood ratio can be defined as

\begin{equation}
\lambda = \frac{\max_{\theta \in H_0}{L(\textbf{X} \mid
\theta)}}{\max_{\theta \in \Theta}{L(\textbf{X} \mid \theta)}} \;.
\label{LikeliratioB}
\end{equation}
This is a statistic because it does not depend on parameters
$\theta$ no more, in the numerator and the denominator there are
likelihood function values at the ML estimators of parameters
$\theta$ with respect to sets $H_0$ and $\Theta$, respectively.

The Wilks's theorem says that under certain regularity conditions if
the hypothesis $H_0$ is true ({\it i.e.} it is true that $\theta \in
H_0$), then the distribution of the statistic $-2\ln{\lambda}$
converges to a $\chi^2$ distribution with $d-k$ degrees of freedom
as $N \longrightarrow \infty$ \cite{James:2006zz,Hoel:1971aa}. The
proof can be found in Ref.~\cite{Dudley:2003ln}. Note that $k = 0$
is possible, so one point in the parameter space (one value of the
parameter) can be tested as well.

\end{document}